\documentclass[12pt,a4paper]{article}

\usepackage{amsmath,amssymb,graphicx,bm}
\usepackage[a4paper,margin=2.7cm]{geometry}
\usepackage{xcolor}

\makeatletter

\def\@jhepkeywords{}
\def\@jhepabstract{}
\def\@jhepauthors{}
\def\@jhepdetails{}

\def\author#1{%
  \ifx\@jhepauthors\@empty
    \gdef\@jhepauthors{#1}%
  \else
    \g@addto@macro\@jhepauthors{, #1}%
  \fi
}

\def\affiliation#1{%
  \g@addto@macro\@jhepdetails{%
    {\itshape\small #1\par}
    \vspace{0.15em}%
  }%
}

\def\emailAdd#1{%
  \g@addto@macro\@jhepdetails{%
    {\small E-mail: \texttt{#1}\par}
    \vspace{0.45em}%
  }%
}

\def\keywords#1{\gdef\@jhepkeywords{#1}}
\def\abstract#1{\gdef\@jhepabstract{#1}}

\def\maketitle{%
  \thispagestyle{plain}%

  \begin{center}

    {\LARGE\bfseries
    \@title\par}

    \vspace{1.0em}

    {\large
    \@jhepauthors\par}

    \vspace{0.6em}

    \@jhepdetails

  \end{center}

  \vspace{0.3em}

  {\noindent\small
  \textbf{Abstract:}\quad
  \@jhepabstract\par}

  \vspace{0.5em}

  {\noindent\small
  \textbf{Keywords:}\quad
  \@jhepkeywords\par}

  \vspace{0.8em}
  \hrule
  \vspace{1.4em}
}

\def\acknowledgments{\section*{Acknowledgments}}

\makeatother

\title{Quantum de Sitter and Analytically-Continued Chern--Simons Theory}

\author{Stephon Alexander}
\author{Kenneth Blakey}

\affiliation{
Brown Center for Theoretical Physics and Innovation,\\
Department of Physics, Brown University,\\
Providence, RI 02912, USA
}

\emailAdd{stephon\_alexander@brown.edu}

\affiliation{
Department of Mathematics, MIT,\\
Cambridge, MA 02139, USA
}

\emailAdd{kblakey@mit.edu}

\abstract{
Insert abstract here.
}


\abstract{We give a Lefschetz thimble definition of the Chern--Simons--Kodama wavefunctional in Lorentzian self-dual gravity with positive cosmological constant. By complexifying the connection and choosing the Lefschetz thimble attached to the self-dual de~Sitter saddle, Witten's analytic continuation of Chern--Simons theory replaces the real contour, on which the defining integral is at best conditionally convergent, by an absolutely convergent integration cycle and yields a controlled semiclassical expansion. The relevant integrand is the full holomorphic exponential $e^{I}$, with $I=\kappa Y_{\rm CS}+(\text{source})$, and the steepest descent flow is governed by the Morse function $h=\operatorname{Re} I$. In the homogeneous isotropic de Sitter reduction, the construction reproduces the Hartle--Hawking Airy wavefunction. In the Bianchi~IX truncation, we take the Lefschetz thimble attached to the expanding de~Sitter saddle (the contracting branch defines a conjugate Lefschetz thimble); within this sector, the symmetric configuration is the only critical point for $\Lambda>0$. The corresponding Lefschetz thimble contains one trace direction with the FRW Airy structure and two anisotropic shear directions that are real-Gaussian damped at leading quadratic order. This gives a finite Bianchi~IX wavefunction which, to leading order, is the FRW Airy factor multiplied by a real anisotropy Gaussian. Thus, the Airy behavior is not an artifact of the one-dimensional minisuperspace reduction, but the leading form of the wavefunction on the Lefschetz thimble. We also discuss how the Lefschetz thimble prescription interfaces with Lorentzian reality conditions.}
\begin{document}
\maketitle
\medskip
\section{Introduction}
The standard $\Lambda$CDM cosmological model remains consistent with current observational data \cite{FreeseToomeyMcDonough}, and a universe dominated at late times by a positive cosmological constant remains one of the simplest and most robust descriptions of the observed universe. Yet, a fully satisfactory quantum description of de Sitter spacetime remains elusive.

In canonical quantum gravity formulated in self-dual variables, there is a distinguished candidate for such a state: the Chern--Simons, or Kodama, wavefunctional. Formally, it is an exact solution of the Gauss, diffeomorphism, and Hamiltonian constraints of self-dual general relativity with positive cosmological constant \cite{Kodama,Smolin,Ashtekar1986,Ashtekar1987,JacobsonSmolin}. Its physical interpretation, however, is obstructed by several well-known difficulties: the relevant Ashtekar connection is $\mathrm{SL}(2,\mathbb{C})$-valued, the naive real integration contour does not define a conventionally normalizable wavefunction, and the implementation of the gravitational reality conditions remains nontrivial. Witten further argued, through comparison with the analogous Yang--Mills state, that perturbations about the Kodama state may contain negative-norm modes and gravitons of negative energy~\cite{Witten2003}. These are genuine physical consistency problems that any viable interpretation of the state must confront.

In this work, we investigate whether a more careful definition of the integration contour, together with the associated reality conditions and semiclassical saddle structure, can clarify the origin of these pathologies and provide new insight into their possible resolution. Two limitations are worth flagging at the outset. First, the analysis below is confined to minisuperspace truncations (FRW and Bianchi~IX); in particular, the perturbative graviton modes about the de~Sitter saddle, the sector in which the negative-energy and negative-norm concerns of \cite{Witten2003} actually arise lie beyond its scope, and are analyzed by complementary methods in \cite{ABKP}. Second, in the Lorentzian theory, the Chern--Simons--Kodama functional transforms by more than a phase under large gauge transformations; this is harmless for defining the contour integrals below, but it matters for interpreting the functional as a state, and we discuss in Secs.~\ref{sec:PL} and~\ref{sec:disc} how the boundary functional embedding of \cite{AlexanderAlexandreDanielssonSpergel} resolves it.

In particular, the purpose of this paper is to give a concrete contour definition of the Lorentzian Kodama state by identifying it with analytically-continued Chern--Simons theory \cite{Witten2010,Witten2010b,Gukov2005,DGLZ,GMP}. The key observation is that Witten's complexified connection has a direct gravitational interpretation in four dimensions. In self-dual gravity, the connection is $A=\Gamma+iK$, where $\Gamma$ is the spatial spin connection and $K$ is the extrinsic curvature. Thus, the complex direction in analytically-continued Chern--Simons theory is not an auxiliary mathematical device when the theory is read gravitationally---it is the direction of extrinsic curvature data. This identification turns analytically-continued Chern--Simons theory into a contour prescription for Lorentzian quantum gravitational initial data.

This Picard-Lefschetz perspective also connects the Kodama state to the familiar sphaleron picture of Chern--Simons theory \cite{AlexanderAlexandreDanielssonSpergel}. In ordinary gauge theory, the Chern--Simons functional organizes configuration space into topological sectors with sphaleron configurations sitting at the top of the barrier between adjacent vacua~\cite{JackiwRebbi,CallanDashenGross,Manton1983,KlinkhamerManton}. In the homogeneous gravitational reduction, the same Chern--Simons functional appears as a function of the isotropic connection variable. The closed de~Sitter saddle is the gravitational analogue of a sphaleron-like configuration in complexified connection space. The unstable direction is not interpreted as tunneling between Yang--Mills vacua, but as the isotropic extrinsic curvature direction whose contour is selected by Picard--Lefschetz theory. This gives the bridge between the Chern--Simons sphaleron intuition and the emergence of the Airy wavefunction in quantum de~Sitter space.

In fact, the sphaleron connection is more than an analogy. Restricting the Plebanski contour prescription of \cite{AlexanderAlexandreDanielssonSpergel} to the Bianchi~IX sector and isolating the boundary Chern--Simons--Kodama functional, as in the Appendix of that work, one finds steepest descent contours that coincide with the Lefschetz thimbles constructed here; the present paper supplies the explicit Picard--Lefschetz theory underlying that choice. This convergence supports a broader interpretation, anticipated by the generalized Hartle--Hawking construction of Herczeg, and Magueijo et al. \cite{AHMHH}: the Chern--Simons--Kodama functional is naturally read as the boundary functional of a generalized, complexified Hartle--Hawking state, with the Lefschetz thimble prescription fixing the contour on which that boundary functional is evaluated. We return to this point in the Discussion.

Once this dictionary is adopted, the Kodama wavefunctional can be treated as a source-deformed analytically-continued Chern--Simons partition function. The source is the variable conjugate to the connection and is therefore identified with the densitized triad. Passing from the connection representation to the metric representation then becomes the corresponding Fourier transform. The essential point is that this transform should not be defined on the naive real contour. Following Witten, we complexify the configuration space and choose the Lefschetz thimble attached to the self-dual de~Sitter saddle. On this steepest descent cycle, the real part of the full holomorphic exponent decreases away from the saddle while its imaginary part is constant. The original oscillatory contour is thereby replaced by an absolutely convergent contour selected by de~Sitter geometry itself.

This construction immediately explains why an Airy function appears in quantum de~Sitter minisuperspace. When the gravitational data are restricted to the homogeneous and isotropic de~Sitter sector, the Chern--Simons functional reduces to a cubic function of the single isotropic extrinsic curvature variable. The metric representation transform is then precisely the Lefschetz thimble version of the Airy integral. Thus, the Hartle--Hawking Airy wavefunction is not imposed as an external minisuperspace ansatz---it emerges from the analytically-continued Chern--Simons--Kodama state after identifying Witten's complex connection with the gravitational pair $(\Gamma,K)$ and evaluating the integral on the de~Sitter truncated Lefschetz thimble. In this sense, the Airy function is the homogeneous isotropic projection of the full Lefschetz thimble.

The main test of this claim is whether the Airy structure survives beyond the one-dimensional reduction. We address this by studying the Bianchi~IX truncation which keeps the isotropic trace mode together with two anisotropic shear modes. This is also the natural setting in which the sphaleron picture becomes useful since the symmetric Bianchi~IX configuration lies on the closed-FRW Chern--Simons profile while the anisotropic directions probe whether the putative saddle is destabilized once the minisuperspace truncation is enlarged. We find that it is not. The trace direction retains the same Airy behavior found in the isotropic model, and the anisotropic directions are Gaussian damped at leading order about the expanding de~Sitter saddle. Equivalently, the leading Bianchi~IX wavefunction is the FRW Airy factor multiplied by a real Gaussian that suppresses anisotropy.

The expanding and contracting de~Sitter branches define conjugate Lefschetz thimbles; the prescription used here selects the expanding branch. Within that branch, the symmetric de~Sitter configuration is the relevant critical point for positive cosmological constant, and the shear modes are exponentially suppressed under the gradient flow. The sphaleron-like Chern--Simons saddle therefore does not lead to an uncontrolled anisotropic instability. Instead, on the Lefschetz thimble, the trace direction produces the Airy wavefunction and the shear directions are damped---this gives a Picard--Lefschetz realization of de~Sitter no-hair in the Bianchi~IX sector when using the upwards gradient flow.

A crucial point throughout the paper is that the contour is governed by the full holomorphic exponent, not by the Chern--Simons phase alone. On the original real slice, one may be tempted to first separate the state into a modulus and a phase and then keep only the oscillatory part. That loses essential information. The Lefschetz thimble leaves the real slice, and, off that slice, the cubic Chern--Simons structure contributes to the steepest descent function; this is exactly what happens in the ordinary Airy integral. The trace mode obtains its Airy behavior from the cubic part of the holomorphic Chern--Simons functional, and the real part of the same holomorphic exponent supplies the Gaussian damping of the shear modes. The Airy behavior, the sphaleron saddle structure, and anisotropy suppression are therefore different aspects of a single contour prescription.

We also explain how this contour prescription interfaces with Lorentzian reality conditions. The Picard--Lefschetz construction defines the holomorphic Chern--Simons integral while the gravitational reality condition yields a holomorphic "coherent-state" inner product \cite{AHF}. At the de~Sitter saddle, the connection has the standard Lorentzian interpretation: $\Gamma$ is the compatible spin connection and $K$ is the de~Sitter extrinsic curvature. Thus, the saddle lies on the expected gravitational reality locus at leading semiclassical order. We do not claim to construct a complete nonperturbative physical inner product for quantum gravity, but we do obtain a controlled semiclassical definition of the Kodama sector around de~Sitter spacetime.

The paper is organized as follows. Section~\ref{sec:setup} reviews the self-dual Ashtekar formulation and identifies the Kodama wavefunctional with the source-deformed analytically-continued Chern--Simons partition function. Section~\ref{sec:PL} formulates the Lefschetz thimble prescription and relates the corresponding downwards gradient flow to the Kapustin--Witten flow in gravitational variables. Section~\ref{sec:dS} identifies the self-dual de~Sitter saddle and shows that the homogeneous isotropic reduction gives the Airy wavefunction. Section~\ref{sec:bianchi} analyzes the Bianchi~IX truncation and shows that the trace mode retains the Airy structure while the shear modes are Gaussian damped. Section~\ref{sec:reality} discusses Lorentzian reality conditions and the AHF inner product. We conclude in Section~\ref{sec:disc} with a discussion and open questions.
\section{The Chern--Simons state from the complexified connection}
\label{sec:setup}
The introduction stated our main perspective: the Chern--Simons wavefunctional, when viewed as a generating functional, is the partition function of Witten's analytically-continued $\mathrm{SL}(2,\mathbb{C})$ Chern--Simons theory. The purpose of this section is to make this perspective precise. We define the holomorphic generating functional $Z[J]$, identify it with the Fourier transform of the Chern--Simons state, fix once and for all the holomorphic exponent $I$ and the Morse function $h=\mathrm{Re}\,I$ we use throughout, and observe that the Ashtekar splitting $\mathcal{A}=\Gamma+iK$ is exactly the splitting $\mathcal{A}=A+i\phi$ Witten uses in the analytically-continued theory (here, the auxiliary field $\phi$ now acquires the gravitational meaning of extrinsic curvature).

\textbf{Self-dual gravity and the canonical pair.}---We first recall the object we view as being analytically continued. Lorentzian general relativity with a positive cosmological constant $\Lambda>0$ can be written, following Ashtekar \cite{Ashtekar1986,Ashtekar1987}, as a theory of a self-dual $\mathrm{SL}(2,\mathbb{C})$ connection whose phase space is coordinatized by the canonical pair
\begin{equation}
\{A^{i}_{a}(x),\,E^{b}_{j}(y)\}\;=\;\ell_{P}^{2}\,\delta^{b}_{a}\,\delta^{i}_{j}\,\delta^{(3)}(x-y),
\label{eq:bracket}
\end{equation}
where: $A^{i}_{a}=\Gamma^{i}_{a}+iK^{i}_{a}$ is the complex self-dual Ashtekar connection on a spatial slice $\Sigma$, with $\Gamma$ and $K$ the real, $\mathfrak{su}(2)$-valued spatial spin connection and extrinsic curvature, respectively; $E^{a}_{i}$ is the conjugate densitized triad; and $\ell_P$ is the Planck length.\footnote{In the self-dual theory, the symplectic structure is often written with an explicit factor of $i$: $\{A,E\}=i\ell_{P}^{2}\,\delta$. We absorb this factor into our conventions so that (1) the triad operator below is $\hat E^{a}_{i}=-\ell_{P}^{2}\,\delta/\delta A^{i}_{a}$ with no factor of $i$ and (2) the Kodama exponent $\kappa$ of Eq. \eqref{eq:Kreview} is real.} Moreover, $i,j,k$ are $\mathfrak{su}(2)$ indices and $a,b,c$ are spatial indices. The theory is pure constraint:
\begin{equation}
G_{i}=D_{a}E^{a}_{i}\approx0,\qquad
V_{a}=E^{b}_{i}F^{i}_{ab}\approx0,\qquad
\mathcal{H}=\epsilon_{ijk}\,E^{a}_{i}E^{b}_{j}\Bigl(F^{k}_{ab}+\tfrac{\Lambda}{3}\,\epsilon_{abc}E^{ck}\Bigr)\approx0;
\label{eq:constraints}
\end{equation}
these are the Gauss, vector (diffeomorphism), and scalar (Hamiltonian) constraints, respectively. Here, $F^{i}_{ab}=\partial_{a}A^{i}_{b}-\partial_{b}A^{i}_{a}+\epsilon^{i}{}_{jk}A^{j}_{a}A^{k}_{b}$ is the curvature and $D_{a}$ is the gauge-covariant derivative. Moreover, observe that $\Lambda$ enters only through the last term $\mathcal{H}$.

In the connection representation, the triad is actually the functional derivative $\hat E^{a}_{i}=-\ell_{P}^{2}\,\delta/\delta A^{i}_{a}$, and the Kodama state is the holomorphic exponential of the Chern--Simons functional:
\begin{equation}
\Psi_{\rm K}[A]=\mathcal{N}\,\exp\!\bigl(\kappa\,Y_{\rm CS}[A]\bigr),\qquad
\kappa=\frac{3}{2\ell_{P}^{2}\Lambda}\in\mathbb{R}_{>0},
\label{eq:Kreview}
\end{equation}
where $\mathcal{N}$ is a formal normalization. Since
\begin{equation}
\frac{\delta Y_{\rm CS}}{\delta A^{i}_{a}}=\epsilon^{abc}F^{i}_{bc}=2B^{a}_{i},
\label{eq:dYCS}
\end{equation}
with $B^{a}_{i}\equiv\tfrac12\epsilon^{abc}F^{i}_{bc}$ the magnetic field, the triad operator acts on \eqref{eq:Kreview} as multiplication by the curvature,
\begin{equation}
\hat E^{a}_{i}\,\Psi_{\rm K}
=-\ell_{P}^{2}\kappa\,\frac{\delta Y_{\rm CS}}{\delta A^{i}_{a}}\,\Psi_{\rm K}
=-\frac{3}{\Lambda}\,B^{a}_{i}\,\Psi_{\rm K} ,
\label{eq:Eeigen}
\end{equation}
i.e.\ $\Psi_{\rm K}$ is the connection representation eigenstate on which the triad is locked to the magnetic field: $E\propto B$.\footnote{The sign is the convention selecting the expanding branch.} Inserting \eqref{eq:Eeigen} for the rightmost triad in $\mathcal{H}$ and using the identity $\epsilon_{abc}B^{ck}=F^{k}_{ab}$,
\begin{equation}
\Bigl(F^{k}_{ab}+\tfrac{\Lambda}{3}\,\epsilon_{abc}\,\hat E^{ck}\Bigr)\Psi_{\rm K}
=\Bigl(F^{k}_{ab}+\tfrac{\Lambda}{3}\,\epsilon_{abc}\bigl(-\tfrac{3}{\Lambda}\bigr)B^{ck}\Bigr)\Psi_{\rm K}
=\bigl(F^{k}_{ab}-F^{k}_{ab}\bigr)\Psi_{\rm K}=0 ,
\label{eq:Hannih}
\end{equation}
shows $\hat{\mathcal{H}}\,\Psi_{\rm K}=0$. Note, the value
$\kappa=3/(2\ell_{P}^{2}\Lambda)$ is exactly what makes the curvature term and the cosmological constant term cancel; moreover, the vanishing bracket in \eqref{eq:Hannih} is the self-dual de~Sitter condition $F^{k}_{ab}=-\tfrac{\Lambda}{3}\epsilon_{abc}E^{ck}$ derived in Sec.~\ref{sec:dS}. The remaining two constraints are immediate: $Y_{\rm CS}$ is diffeomorphism invariant and gauge invariant up to the large gauge shifts of Sec.~\ref{sec:PL}, so $\Psi_{\rm K}$ is annihilated by $\hat G_{i}$ and $\hat V_{a}$ as well. The Kodama state is thus an exact solution of all the constraints of self-dual gravity with $\Lambda>0$; the rest of this section recasts that solution as an analytically-continued Chern--Simons generating
functional.

\textbf{Analytically-continued Chern--Simons theory.}---
We start from Witten's analytic continuation of Chern--Simons theory
\cite{Witten2010,Witten2010b}. Here, the connection is a holomorphic field $\mathcal{A}$ coming from a complex configuration space $\mathcal{A}_\mathbb{C}$, and the partition function is defined by integration along a chosen contour $\mathcal{C}\subset\mathcal{A}_{\mathbb{C}}$. Observe, in the gravitational case, the self-dual connection $\mathcal{A}=\Gamma+iK$ is already $\mathrm{SL}(2,\mathbb{C})$-valued, so $\mathcal{A}_{\mathbb{C}}$ is its natural home, while the Lorentzian theory sits on the real slice.

The Chern--Simons functional on a Cauchy slice $\Sigma$ is
\begin{equation}
Y_{\rm CS}[\mathcal{A}] \;=\; \int_{\Sigma}\!\mathrm{tr}\!\bigl(
\mathcal{A}\wedge d\mathcal{A}
+ \tfrac{2}{3}\,\mathcal{A}\wedge\mathcal{A}\wedge\mathcal{A}
\bigr) .
\label{eq:YCS}
\end{equation}
We define the \emph{holomorphic exponent}
\begin{equation}
I[\mathcal{A};J] \;\equiv\; \kappa\,Y_{\rm CS}[\mathcal{A}]
\;+\;\!\int_{\Sigma}\!J\!\cdot\!\mathcal{A} ,
\label{eq:Idef}
\end{equation}
with $J^{a}_{i}$ a linear source coupled to the connection, and the \emph{source-deformed partition function}
\begin{equation}
Z[J] \;=\; \int_{\mathcal{C}}\!\mathcal{D}\mathcal{A}\;
\exp\!\bigl(\,I[\mathcal{A};J]\,\bigr) .
\label{eq:ZJ}
\end{equation}
As a result, the integrand $e^{I}$ is the full holomorphic exponential, and the
Picard--Lefschetz Morse function—the ``height'' function whose
downward gradient flow will define the Lefschetz thimbles—is
\begin{equation}
h\;\equiv\;\mathrm{Re}\,I.
\label{eq:Morse}
\end{equation}
Note, on the original real cycle, the integrand of \eqref{eq:ZJ} oscillates and the integral is at best conditionally convergent, exactly as in ordinary Chern--Simons theory. In particular, Picard--Lefschetz theory circumvents this by allowing us to replace the original conditionally convergent contour with an absolutely convergent contour via utilizing Lefschetz thimbles; this is explained in Sec.~\ref{sec:PL}.

We now show that this partition function is the Chern--Simons--Kodama state in the metric representation, obtained by Fourier transforming its connection representation form. Identifying
$\mathcal{A}$ with the complex self-dual $\mathrm{SL}(2,\mathbb{C})$ Ashtekar
connection on $\Sigma$, the Chern--Simons wavefunctional in the connection
representation is
\begin{equation}
\Psi_{\rm K}[\mathcal{A}]
\;=\;
\mathcal{N}\,\exp\!\bigl(\kappa\,Y_{\rm CS}[\mathcal{A}]\bigr).
\label{eq:Kodama}
\end{equation}
To pass to the metric representation we use the canonical structure of the
self-dual variables. The connection $\mathcal{A}$ and the densitized triad $E$ are
conjugate, so, in the connection representation, $\hat E$ acts as a functional
derivative in $\mathcal{A}$. The self-dual reality structure removes the usual
factor of $i$, so this action is holomorphic:
\begin{equation}
\bigl[\hat{\mathcal{A}}^{a}_{i}(x),\,\hat E^{b}_{j}(y)\bigr]
=\ell_{P}^{2}\,\delta^{ab}\delta_{ij}\,\delta^{(3)}(x,y),
\qquad
\hat E^{a}_{i}=-\,\ell_{P}^{2}\,\frac{\delta}{\delta\mathcal{A}^{a}_{i}} .
\label{eq:CCR}
\end{equation}
The eigenfunctionals of $\hat E$ that follow from \eqref{eq:CCR} are the
exponential kernels $\exp\!\bigl(\ell_{P}^{-2}\!\int_{\Sigma}E\cdot\mathcal{A}\bigr)$, with eigenvalue $E$, and the change from the connection to the metric representation is the functional transform of $\Psi_{\rm K}[\mathcal{A}]$ against this kernel:
\begin{equation}
\Psi_{\rm K}[E]
=\;
\mathcal{N}\!\int_{\mathcal{C}}\!\mathcal{D}\mathcal{A}\;
\exp\!\Bigl(\kappa\,Y_{\rm CS}[\mathcal{A}]
+\ell_{P}^{-2}\!\int_{\Sigma}E\cdot\mathcal{A}\Bigr).
\label{eq:Psi-E}
\end{equation}
Two remarks make this transform well-defined and fix its meaning. First, since $Y_{\rm CS}$ is holomorphic, the integral is only conditionally convergent on the real slice, so $\int_{\mathcal{C}}$ is not a number until a suitable contour $\mathcal{C}\subset\mathcal{A}_{\mathbb{C}}$ is specified; that choice is the Picard--Lefschetz problem of Sec.~\ref{sec:PL}, and it is the cycle---not the integrand alone---that determines which solution of the constraints $\Psi_{\rm K}[E]$ is. Second, as we will soon see, the kernel of \eqref{eq:CCR} becomes the oscillatory Fourier kernel $\exp\!\bigl(+iV_{c}\,p\!\cdot\!b/\ell_{P}^{2}\bigr)$ used in Sec.~\ref{sec:bianchi}, with $b=\mathrm{Im}\,\mathcal{A}$ the real extrinsic curvature variable; its sign, or equivalently the orientation of $\mathcal{C}$, is the one that places the saddle on the expanding de~Sitter branch.

With this understood, \eqref{eq:Psi-E} is precisely Witten's source construction, with \(J\) identified as the triad conjugate to \(\mathcal{A}\). Evaluating \(Z[J]\) on a cycle in \(\mathcal{A}_{\mathbb{C}}\) then yields the Chern--Simons state in the metric representation on that same cycle. This is just a change of
representation, and the only input is the choice of $\mathcal{C}$.

Moreover, the Ashtekar variable splits the holomorphic connection as
\begin{equation}
\mathcal{A}^{i}_{a} \;=\; \Gamma^{i}_{a}(E) \;+\; i\,K^{i}_{a} ,
\label{eq:dictionary}
\end{equation}
with $\Gamma$ and $K$ the spin connection and extrinsic curvature, respectively. This is precisely the splitting Witten uses for the analytically-continued theory. Under the dictionary $A_{\rm Witten}=\Gamma$, $\phi_{\rm Witten}=K$, the auxiliary Higgs-like field $\phi$ acquires the gravitational meaning of extrinsic curvature, and each point of $\mathcal{A}_{\mathbb{C}}$ is read as a piece of spacetime Cauchy data $(E,K)$.

We refer to the locus of Cauchy data at fixed $E$,
\begin{equation}
\mathcal{F}_{E}\;\equiv\;\bigl\{\,\Gamma(E)+iK\;:\;K\ \text{real and $\mathfrak{s}\mathfrak{u}(2)$-valued}\,\bigr\}\subset\mathcal{A}_{\mathbb{C}},
\label{eq:fibre}
\end{equation}
as the \emph{Lorentzian fibre}. Note, this is the original integration contour of the transform \eqref{eq:Psi-E}. On $\mathcal{F}_{E}$, the kernel factorizes as $\exp\!\bigl(\ell_{P}^{-2}\!\int E\!\cdot\!\Gamma(E)\bigr)\,\exp\!\bigl(i\,\ell_{P}^{-2}\!\int E\!\cdot\!K\bigr)$. In particular, the first factor does not depend on the integration variable and is absorbed into the normalization of $\Psi_{\rm K}[E]$; hence, on the fibre, the effective holomorphic exponent is $I=\kappa\,Y_{\rm CS}+i\,\ell_{P}^{-2}\!\int E\!\cdot\!K$, continued holomorphically in $K$. Since the absorbed factor is constant on $\mathcal{A}_{\mathbb{C}}$ at fixed $E$, it does not affect: saddles, Hessians, or the gradient flow--- we use the two forms of $I$ interchangeably.

Via the splitting \eqref{eq:dictionary}, the holomorphic functional decomposes as $Y_{\rm CS} = \mathrm{Re}\,Y_{\rm CS} + i\,\mathrm{Im}\,Y_{\rm CS}$, so the wavefunctional \eqref{eq:Kodama} factorizes into a modulus and a phase:
\begin{equation}
\Psi_{\rm K} \;=\; \mathcal{N}\,
\underbrace{\exp\!\bigl(\kappa\,\mathrm{Re}\,Y_{\rm CS}\bigr)}_{\text{modulus}}\,
\underbrace{\exp\!\bigl(i\,\kappa\,\mathrm{Im}\,Y_{\rm CS}\bigr)}_{\text{phase}}.
\label{eq:phase-modulus}
\end{equation}
This factorization plays no role in the contour construction. Picard--Lefschetz theory is applied to the full holomorphic exponent $I=\kappa Y_{\rm CS}+(\text{source})$ through its Morse function $h=\mathrm{Re}\,I$ on $\mathcal{A}_{\mathbb{C}}$, not to the modulus $\kappa\,\mathrm{Re}\,Y_{\rm CS}$ alone. The two agree only on the Lorentzian fibre \eqref{eq:fibre} where the source coupling is purely imaginary:
\begin{equation}
\mathrm{Re}\,I\big|_{\mathcal{F}_{E}} \;=\; \kappa\,\mathrm{Re}\,Y_{\rm CS}\big|_{\mathcal{F}_{E}};
\label{eq:onaxis}
\end{equation}
this restriction is quadratic in the extrinsic curvature and contains no cubic. The Lefschetz thimble, however, leaves the fibre, and off it the holomorphic cubic of $Y_{\rm CS}$ contributes to $\mathrm{Re}\,I$---exactly as the canonical Airy exponent $I=i(t^{3}/3+xt)$ has $\mathrm{Re}\,I=-\mathrm{Im}(z^{3}/3+xz)$ off the real axis. Retaining the full $I$ is therefore what produces both the trace-mode Airy behavior and the \emph{real} Gaussian damping of the shear modes in Sec.~\ref{sec:bianchi}; dropping the modulus and keeping only the phase loses the latter (Appendix~\ref{app:fullmorse}, with the bookkeeping identity $\mathrm{Re}\,\tilde I=-\mathrm{Im}\,S$ of the pure phase reduction recorded in Appendix~\ref{app:phasechoice}). The modulus reappears independently as the intrinsic-curvature weight of the AHF norm in Sec.~\ref{sec:reality}, where, at the de~Sitter saddle, it equals the finite Chern--Simons invariant $Y_{\rm CS}(\Gamma)=\mathrm{Re}\,Y_{\rm CS}(\mathcal{A}_{*})$.
\section{Connecting Lefschetz thimbles to the
         Kapustin--Witten equation}
\label{sec:PL}
The previous section contained the observation that $Z[J]$, on the original real slice, is at best conditionally convergent. The purpose of this section is to formulate the choice of integration cycle properly and give it a physical interpretation via Picard--Lefschetz theory. In particular, we will take a Lefschetz thimble itself as the defininig cycle. Moreover, we identify the downwards gradient flow that sweeps out this Lefschetz thimble with the Kapustin--Witten equation in gravitational variables.

A well-defined formulation of the contour used in \eqref{eq:ZJ} is provided by Picard--Lefschetz theory \cite{Witten2010,Witten2010b}. The critical points of the holomorphic exponent $I[\mathcal{A};J]$ of \eqref{eq:Idef}, i.e.\ the points solving $\delta I/\delta\mathcal{A}=0$, are the saddle configurations $\mathcal{A}_\alpha$. We treat the real-valued function of \eqref{eq:Morse} as a \emph{Morse function} on $\mathcal{A}_{\mathbb{C}}$, i.e. a function whose critical points are \emph{non-degenerate} (invertible Hessian). Strictly speaking, $h$ is a Morse function only when all of its critical points are nondegenerate. This condition fails on the \emph{discriminant locus} where two or more critical points merge and the Hessian of $h$ develops a zero mode; the turning point ($\bar{P}=0)$ discussed in Sec.~5 is the corresponding minisuperspace example. Meanwhile, \emph{Stokes loci/walls} are distinct, where the critical points remain nondegenerate, but their imaginary critical values align, causing the associated Lefschetz thimble decomposition (defined below) to jump. In the full field theory, the Morse property must furthermore be understood after gauge fixing, or equivalently on the quotient by gauge transformations and spatial diffeomorphisms, since the corresponding symmetry directions generate zero modes of the Hessian. Throughout, we assume generic parameters away from both the discriminant and Stokes loci.

Each saddle $\mathcal{A}_{\alpha}$ then carries two submanifolds of $\mathcal{A}_{\mathbb{C}}$: the Lefschetz \emph{thimble} $\mathcal{J}_{\alpha}$, swept out by the downwards gradient flow, and the dual Lefschetz \emph{anti-thimble} $\mathcal{K}_{\alpha}$, swept out by the upwards gradient flow. Along $\mathcal{J}_{\alpha}$, the imaginary part $\mathrm{Im}\,I$ is constant while $\mathrm{Re}\,I$ decreases monotonically away from the saddle, so the integrand $e^{I}$ decays exponentially. Thus, any integral over a Lefschetz thimble is absolutely convergent. Moreover, a contour on which an integral is absolutely (or conditionally) convergent admits an integer-weighted decomposition
\begin{equation}
\mathcal{C} \;=\; \sum_{\alpha} n_{\alpha}\,\mathcal{J}_{\alpha},
\qquad
n_{\alpha} \;=\; \bigl\langle\mathcal{C},\,\mathcal{K}_{\alpha}\bigr\rangle
\in \mathbb{Z},
\label{eq:thimble-decomp}
\end{equation}
where the integers $n_{\alpha}$ are intersection numbers of the original cycle with the dual anti-thimbles; these integers jump discontinuously across Stokes walls. In particular, the selection of any single $\mathcal{J}_\alpha$, together with the weight $n_\alpha=1$, amounts to a definite homotopy class of contour.

The following two remarks should orient the reader in regards to the Lefschetz decomposition. First, as already mentioned, Picard--Lefschetz theory presupposes a convergent integration cycle to apply the decomposition \eqref{eq:thimble-decomp} to. In particular, this is the situation in every explicit computation of this paper. Second, for the formal functional integral \eqref{eq:ZJ}, no convergent cycle is given in advance; i.e. we must provide a choice of cycle, so this is a part of the definition of the state. In this paper, we do this by choosing a suitable Lefschetz thimble $\mathcal{C}=\mathcal{J}_*$, so $n_\alpha=\delta_{\alpha*}$.\footnote{However, in the classically allowed region of the FRW truncation below, we will instead choose the Lefschetz thimbles through both the de Sitter saddle and its conjugate (both with unit weight) in order to make the comparison to canonical Airy.\label{footnote}} Moreover, for any allowable contour, integration by parts produces no boundary terms, so the resulting integral solves the same constraint equations; the contour labels which solution of the constraints the state is---exactly as anticipated below \eqref{eq:Psi-E}. Statements in this paper that the Lefschetz thimble ``replaces'' the real slice are to be read in this sense. In particular, we would like to emphasize that we do not compute the Lefschetz decomposition of the Lorentzian fibre \eqref{eq:fibre}.

With this understood, the intersection machinery of \eqref{eq:thimble-decomp} plays two distinct roles here. In the full construction, it is nearly trivial by design since, as already mentioned, we define the state by $\mathcal{C}=\mathcal{J}_{*}$, so $n_{\alpha}=\delta_{\alpha *}$, and the single unit weight factorizes out of every subsequent calculation (the only exception is when we choose a sum of Lefschetz thimbles with unit weight in the classically allowed region of the FRW truncation below, cf. Footnote \ref{footnote}). Meanwhile, the intersection numbers must be computed precisely when a contour is specified in advance.

\textbf{Gradient flow as Kapustin--Witten equation.}---As already mentioned, the flow that sweeps out $\mathcal{J}_{\alpha}$ is the downwards gradient flow of $\mathrm{Re}\,I$ on the complex configuration space $\mathcal{A}_\mathbb{C}$; this is expressed in field-theoretic form as
\begin{equation}
\frac{d\mathcal{A}_{i}}{ds}
\;=\;
-\,\overline{\frac{\delta\bigl(\mathrm{Re}\,I[\mathcal{A};J]\bigr)}{\delta\mathcal{A}_{i}}}
\ ,
\label{eq:PL-grad}
\end{equation}
where $\mathcal{A}_{i}(x)$ is the spatial complex connection on $\Sigma$ and the bar denotes complex conjugation. In particular, \eqref{eq:PL-grad} is the downwards gradient flow of $\mathrm{Re}\,I$ for the flat $L^{2}$ K\"ahler metric of $\mathcal{A}_{\mathbb{C}}$ (Appendix~\ref{app:gradflow}). Under the identification of Witten's parameterization of the $\rm{SL}(2,\mathbb{C})$ connection with the Lorentzian $\mathrm{SL}(2,\mathbb{C})$ Ashtekar connection, \eqref{eq:dictionary} turns \eqref{eq:PL-grad} on $\Sigma\times\mathbb{R}_{+}$, in temporal gauge, into the Kapustin--Witten equation \cite{KapustinWitten,Witten1101} expressed in gravitational variables,
\begin{align}
F[\Gamma]\,-\,K\wedge K\,+\,\tau\,d_{\Gamma} K &\;=\;0 ,
\nonumber\\
d_{\Gamma}\!\star\!K &\;=\;0 ,
\label{eq:KW}
\end{align}
with $\tau$ a suitable parameter \cite{Witten2010} which equals $\kappa$ in this paper and the saddle $(\Gamma_*,K_*)$ as boundary condition at $s\to-\infty$. Note, \eqref{eq:KW} is written for the undeformed functional ($J=0$). Meanwhile, the source adds the constant inhomogeneous term $-\bar J$ to the flow (cf.\ Appendix~\ref{app:gradflow}, Eq.~\eqref{eq:app-flow-explicit}); this shifts the fixed points from flat connections to the saddles of the source-deformed exponent but does not alter the character of the equations. In particular, the Lefschetz thimble $\mathcal{J}_*$ is the moduli space of solutions $(\Gamma(s),K(s))$ on $\Sigma\times\mathbb{R}_{+}$ obeying \eqref{eq:KW} with that boundary condition. In the next section, $\mathcal{J}_{*}$ will be the descending (or unstable) manifold of the de~Sitter saddle under the downwards gradient flow: again, it is parametrized by spatial Cauchy data $(E_{0},K_{0})$ on $\Sigma$, emanating from $(\Gamma_{\rm dS},K_{\rm dS})$, modulo $\mathrm{SL}(2,\mathbb{C})$ gauge and spatial diffeomorphisms; equivalently, $(\Gamma_{\rm dS},K_{\rm dS})$ is the attractor of the time-reversed (upwards gradient) flow.

The dictionary rephrases the Witten construction into a 4D Lorentzian, self-dual gravitational framework. In pure analytically-continued Chern--Simons theory, the auxiliary $\mathfrak{su}(2)$-valued 1-form $\phi$ supplies the imaginary direction needed for analytic continuation but has no \emph{a priori} physical content. The identification of the auxiliary field with the extrinsic curvature, $\phi_{\rm Witten}=K$, assigns it the gravitational meaning of extrinsic curvature. Moreover, each point of $\mathcal{J}_{*}$ becomes a piece of spacetime Cauchy data, the de~Sitter saddle becomes a maximally symmetric slice with its embedding data, and the Picard--Lefschetz manifestation of cosmic no-hair derived in Sec.~\ref{sec:bianchi} becomes a Picard--Lefschetz attractor theorem on this physical Cauchy data space.\footnote{For the present application, $\kappa = 3/(2\ell_{P}^{2}\Lambda)$ is real and positive for $\Lambda > 0$. In Witten's complex-$\tau$ parametrized moduli space of analytically-continued Chern--Simons theories \cite{Witten2010,Witten1101}, this value sits generically away from Stokes walls, so the integers $n_{\alpha}$ in \eqref{eq:thimble-decomp} are constant in a neighborhood of $\kappa$ in the physical regime $\Lambda\,\ell_{P}^{2}\ll 1$. Also, the gauge quotient throughout is taken with respect to the identity component of $\mathrm{SL}(2,\mathbb{C})$; topologically nontrivial (``large'') gauge transformations shift $\kappa\,Y_{\rm CS}$ by integer multiples of $2\pi\kappa$, hence generate multi-valuedness for non-integer $\kappa$ that is absorbed on the universal cover of the gauge orbit space~\cite{Witten2010,Kodama,Smolin}.\label{footnote:gaugeorbit}}

One point in Footnote \ref{footnote:gaugeorbit} deserves promotion to the text. Passing to the universal cover of the gauge orbit space makes the Lefschetz thimble integrals well-defined, but it does not restore gauge invariance of the state itself: in the Lorentzian theory, $\kappa$ is real, so, under a large gauge transformation of winding number $n$, the exponent $\kappa\,Y_{\rm CS}$ shifts by the \emph{real} quantity $2\pi\kappa n$ and the Chern--Simons--Kodama functional is rescaled by $e^{2\pi\kappa n}$---it transforms by more than a phase. The contour integrals of this paper are insensitive to this since each Lefschetz thimble lies in a fixed topological sector and the decomposition \eqref{eq:thimble-decomp} is computed at a fixed lift. However, any norm built from $|\Psi_{\rm K}|^{2}$, including the AHF norm of Sec.~\ref{sec:reality}, then fails to be invariant under large gauge transformations; this is no obstacle to defining the integrals, but it does obstruct interpreting the functional \emph{by itself} as a physical state on the gauge quotient. The resolution we adopt is that of \cite{AlexanderAlexandreDanielssonSpergel}: the Chern--Simons--Kodama functional should be understood as a boundary functional embedded in a gauge-invariant state of the Plebanski path integral so that the gauge variance of the boundary term is compensated by the bulk into which it embeds, in the manner of anomaly inflow. On this reading, the Lefschetz thimble construction of this paper defines the boundary functional and fixes its integration cycle while the large gauge dependence is absorbed by the nonperturbative completion. We return to this in Sec.~\ref{sec:disc}.
\section{de~Sitter saddle and the Hartle-Hawking Airy wavefunction}
\label{sec:dS}
With the Picard--Lefschetz contour selection now in place, this section identifies the relevant saddle as the self-dual de~Sitter geometry, and tests the resulting construction in the simplest reduction beyond the abstract setup. Restricting the contour integral to the homogeneous isotropic minisuperspace collapses the trace direction to a single cubic variable, and the construction reproduces the Airy wavefunction obtained for de~Sitter quantum cosmology in \cite{Magueijo,Randono}. This 1D template fixes the analytic structure that the next section will test against anisotropy.

The critical point equation for the holomorphic exponent \eqref{eq:Idef} reads
\begin{equation}
\kappa\,\epsilon^{abc}F^{i}_{bc}[\mathcal{A}]
\;+\; J^{ai} \;=\; 0;
\label{eq:saddle}
\end{equation}
this is the stationary (i.e. $\partial_{s}=0$) version of the Kapustin--Witten equation \eqref{eq:KW}: an $s$-independent solution of \eqref{eq:KW} is automatically a solution of \eqref{eq:saddle} and vice versa. In particular, we may take the self-dual de~Sitter geometry,
\begin{equation}
F^{i}_{ab}[\mathcal{A}_{*}]
\;=\;
-\,\frac{\Lambda}{3}\,\epsilon_{abc}\,E^{ci} ,
\label{eq:dS}
\end{equation}
solved by the Ashtekar connection of de~Sitter geometry,
\begin{equation}
\mathcal{A}_{*}^{i}\;=\;\Gamma^{i}(E)\;+\;i\,K_{\rm dS}^{i}(E),
\quad
K_{\rm dS}^{i}\;=\;\sqrt{\Lambda/3}\,e^{i},
 ,
\label{eq:saddleconn}
\end{equation}
where $e^{i}$ is the spatial cotriad compatible with $E^{a}_{i}$ and $\Gamma^{i}(E)$ the corresponding torsion-free spin connection, as the asymptotic condition to define the Lefschetz thimble from Sec.~\ref{sec:PL}.

The factor of $i$ in $\mathcal{A}_{*}=\Gamma+iK_{\rm dS}$ is the explicit self-dual Ashtekar prescription. Substituting \eqref{eq:dS} into \eqref{eq:saddle} and using $\epsilon^{abc}\epsilon_{bcd}=2\delta^{a}_{d}$ pins the source as
\begin{equation}
J^{ai} \;=\; \frac{2\kappa\Lambda}{3}\,E^{ai}\;=\;\frac{1}{\ell_{P}^{2}}\,E^{ai};
\label{eq:JE}
\end{equation}
so, the Fourier kernel $\exp(\int J\!\cdot\!\mathcal{A})$ in \eqref{eq:ZJ} is precisely the canonical kernel of \eqref{eq:Psi-E}, i.e.\ the Schr\"odinger-style transform from the connection to the metric representation. The source coupling $\int J\!\cdot\!\mathcal{A}$ varies under $\mathrm{SL}(2,\mathbb{C})$ gauge transformations by a term proportional to $d_{\mathcal{A}}J$; with the identification \eqref{eq:JE}, this is the Gauss constraint $D_{a}E^{a}_{i}=0$, which holds on the constraint surface, so the saddle equation is gauge-covariant on-shell.

We define the Kodama generating functional as the contour integral on the Lefschetz thimble through $\mathcal{A}_{*}$:
\begin{equation}
Z_{\rm K}[J]\;\equiv\;\int_{\mathcal{J}_{*}}\!\mathcal{D}\mathcal{A}\,
\exp\!\bigl(I[\mathcal{A};J]\bigr) .
\label{eq:ZK}
\end{equation}
In particular, as already mentioned, the Lorentzian fibre \eqref{eq:fibre} is the original integration contour of the transform \eqref{eq:Psi-E}; on it, however, the transform is defined at best as a conditionally convergent integral (and, in the full theory, only formally). Following the logic of Sec.~\ref{sec:PL}, we therefore define \eqref{eq:ZK} on the allowed cycle given by the Lefschetz thimble through the de~Sitter saddle (with unit weight in terms of the decomposition~\eqref{eq:thimble-decomp}); in the reductions below, where the fibre integral is conditionally convergent, this choice is related to it by \eqref{eq:thimble-decomp} with the intersection numbers stated there.

\textbf{Isotropic reduction to a single ordinary integral.}---The canonical Airy
integral
\begin{equation}
\mathrm{Ai}(x) \;=\; \frac{1}{2\pi}\!\int_{-\infty}^{\infty}\!\!dt\,
\exp\!\bigl[\,i\bigl(\tfrac{1}{3}t^{3} + x\,t\bigr)\bigr],
\label{eq:Airy}
\end{equation}
on the real axis, has a purely oscillatory integrand and converges only conditionally. Following \cite{Witten2010,Witten2010b}, to make it converge, complexify $t\to z\in\mathbb{C}$ and study the saddles $z_{\pm}^{2}=-x$. For $x>0$ the real axis is homotopic, with no Stokes obstruction, to the single Lefschetz thimble through $z_{+}=i\sqrt{x}$, and, as $x\to+\infty$, the integral has the standard exponential asymptotic $\mathrm{Ai}(x)\sim (1/2\sqrt{\pi}\,x^{1/4})\,\exp(-\tfrac{2}{3}x^{3/2})$ with multiplicities $(n_{+},n_{-})=(1,0)$. For $x<0$ both real saddles contribute, $(n_{+},n_{-})=(1,1)$, and, as $x\to-\infty$, the result oscillates: $\mathrm{Ai}(x)\sim (1/\sqrt{\pi}\,|x|^{1/4})\,\sin(\tfrac{2}{3}|x|^{3/2}+\tfrac{\pi}{4})$. The discontinuous change at $x=0$ is the prototypical Stokes phenomenon.

We now show that this Airy structure arises from the Chern--Simons--Kodama partition function (\ref{eq:ZK}). To exhibit the construction concretely, we consider a flat slicing of a homogeneous isotropic (FRW) connection within a fiducial cell of finite comoving volume $V_0$: $\mathcal{A}^{i}_{a}=c\,\delta^{i}_{a}$, $E^{a}_{i}=p\,\delta^{a}_{i}$.\footnote{The computation would work just as well in the closed slicing; we use the flat slicing for brevity.} The real configuration variable is the homogeneous extrinsic curvature $b\equiv\mathrm{Im}\,\mathcal{A}$. In the flat slicing the spin connection vanishes, so the connection is purely the imaginary direction, thus the modulus is absent and we are dealing with a pure phase. On this ansatz $Y_{\rm CS}=-2iV_{0}b^{3}$, so $\Psi_{\rm K}(b)=\mathcal{N}\,\exp(-2i\kappa V_{0}b^{3})$, and Fourier transforming to the metric representation \eqref{eq:Psi-E} yields the single ordinary integral
\begin{equation}
Z_{\rm K}^{\rm MS}(p)
\;=\;
\int_{\mathcal{J}^{\rm MS}_{*}}\!db\,
\exp\!\Bigl[-\,i\,\frac{3V_{0}}{\ell_{P}^{2}}\Bigl(\frac{b^{3}}{\Lambda}-p\,b\Bigr)\Bigr] ;
\label{eq:ZKMS}
\end{equation}
on the original contour, the integral was conditionally convergent, but it is now rendered absolutely convergent by either (1) the FRW truncation of the Lefschetz thimble through the de Sitter saddle when $p<0$ or (2) the FRW truncation of the Lefschetz thimble through the de Sitter saddle together with its truncated conjugate when $p>0$, both contours denoted $\mathcal{J}^{\rm MS}_{*}$.

Eq.~\eqref{eq:ZKMS} is essentially already in canonical Airy form. Rescaling $b=-\eta\,z$ with $\eta^{3}\equiv\ell_{P}^{2}\Lambda/(9V_{0})$ canonicalizes the cubic, and
\begin{equation}
Z_{\rm K}^{\rm MS}(p)
\;=\;\eta\!\int_{\mathcal{J}^{\rm MS}_{*}}\!dz\,
\exp\!\bigl[\,i\bigl(\tfrac{1}{3}z^{3}+x\,z\bigr)\bigr]
\;=\;2\pi\,\eta\,\mathrm{Ai}(x),
\qquad
x\;=\;-\,\frac{3V_{0}\,p}{\ell_{P}^{2}}\,\eta ,
\label{eq:KodamaAiry}
\end{equation}
with $x$ real and, for physical $p>0$, negative. The saddles $b=\pm\sqrt{\Lambda p/3}$ are the real expanding and contracting de~Sitter branches, the flat slicing has no turning point, and the wavefunction oscillates throughout, as appropriate for Lorentzian de~Sitter (the closed slicing restores the turning point through the spatial curvature term, cf.\ $x_{\rm IX}$ in Sec.~\ref{sec:bianchi}). Observe that $Z_{\rm K}^{\rm MS}(p)$ exhibits Hartle--Hawking-type behavior~\cite{HartleHawking}: $\mathrm{Ai}(x)$ is real and exponentially decaying for $x>0$ (classically forbidden) and oscillatory for $x<0$ (classically allowed). Moreover, for $x<0$ both Lefschetz thimbles carry unit weight, $(n_{+},n_{-})=(1,1)$, and their sum is the real, oscillatory $\mathrm{Ai}(x)$, whereas retaining a single Lefschetz thimble isolates one WKB branch---the Vilenkin tunneling combination $\mathrm{Ai}\mp i\,\mathrm{Bi}$~\cite{Vilenkin}. The real contour computation of this FRW transform of the Chern--Simons state, exhibiting precisely this Airy/Hartle--Hawking--Vilenkin structure, is due to Magueijo \cite{Magueijo} (see also \cite{AHMHH}); in the present language, Magueijo's real-$b$ contour is the $(n_{+},n_{-})=(1,1)$ member of the decomposition \eqref{eq:thimble-decomp}. For Picard--Lefschetz analyses of Lorentzian quantum cosmology see also \cite{FLT}.\footnote{The 1D Airy truncated Lefschetz thimble in \eqref{eq:KodamaAiry} is not grafted onto the full theory: it is the restriction of the infinite-dimensional Lefschetz thimble $\mathcal{J}_{*}\subset\mathcal{A}_{\mathbb{C}}$ through the de~Sitter saddle to the FRW minisuperspace submanifold $\mathcal{M}_{\rm MS}\equiv\{\mathcal{A}^{i}_{a}=c\delta^{i}_{a}: c\in\mathbb{C}\}$. Restricting \eqref{eq:ZJ} to $\mathcal{M}_{\rm MS}$ produces exactly \eqref{eq:ZKMS}, the de~Sitter Ashtekar connection $\mathcal{A}_{*}$ restricts to the expanding branch saddle $c_{+}$, and the Kapustin--Witten equation \eqref{eq:KW} reduces on $\mathcal{M}_{\rm MS}$ to the steepest descent ODE $db/ds = -\overline{\partial_{b}I^{\rm MS}}$ with $I^{\rm MS}(b)\equiv -i\tfrac{3V_{0}}{\ell_{P}^{2}}(b^{3}/\Lambda-pb)$. The FRW Airy wavefunction is therefore the homogeneous isotropic projection of the full Picard--Lefschetz construction.}
\section{Bianchi IX: Anisotropic test of Airy dominance}
\label{sec:bianchi}
The homogeneous isotropic de Sitter reduction presented in Sec.~\ref{sec:dS} produced an Airy wavefunction, but only after collapsing the connection to a single isotropic variable. The natural worry is that this Airy form is an artifact of that 1D collapse and would not survive once anisotropy is restored. The Bianchi~IX truncation is the next simplest reduction to test this as it keeps the isotropic (volume) mode together with two anisotropic (shear) modes. In the metric representation the Kodama state is the contour integral $Z_{\rm K}^{\rm IX}[p]=\int\exp(I^{\rm IX})\,d^{3}b$ over the real extrinsic curvature $b_{i}$, with $I^{\rm IX}=\kappa Y_{\rm CS}^{\rm IX}+(\text{source})$ the full holomorphic exponent. On the real-$b$ contour, this integrand is oscillatory and the integral only conditionally convergent, so we exchange it for the Bianchi~IX truncated Lefschetz thimble through the de~Sitter saddle, the descending manifold of the Morse function $h^{\rm IX}=\mathrm{Re}\,I^{\rm IX}$, where it converges absolutely.

We would like to emphasize that $I^{\rm IX}$ must be kept in full. Splitting it as $I^{\rm IX}=\kappa\,\mathrm{Re}\,Y_{\rm CS}^{\rm IX}+iS^{\rm IX}$ separates a real modulus $\kappa\,\mathrm{Re}\,Y_{\rm CS}^{\rm IX}$---quadratic in $b$ and nonzero only because the spin connection does not vanish on the closed slicing---from an oscillatory phase $S^{\rm IX}$ which carries the cubic. Neither piece may be dropped since the two control opposite directions. Along the two shear directions, the modulus is already a decaying Gaussian on the real-$b$ contour, so anisotropy is exponentially suppressed and those integrals converge with no contour change at all. Meanwhile, along the trace direction, the modulus is flat and the integrand becomes a purely oscillatory cubic. This is the only conditionally convergent sector and therefore the only one that actually requires a Lefschetz thimble, yielding the canonical Airy integral. The resulting factorization identifies the trace contribution with the FRW Airy function of Sec.~\ref{sec:dS} mode for mode, while the two shear directions contribute a finite, real Gaussian that suppresses anisotropy; this demonstrates that the Airy form is not an artifact of the 1D collapse. The rest of this section makes each step precise.

\textbf{Setup and ansatz.}---The anisotropic Bianchi~IX metric is
\begin{equation}
ds^{2} \;=\; -N^{2}(t)\,dt^{2} \;+\; a_{i}^{2}(t)\,(\omega^{i})^{2} ,
\label{eq:IX-metric}
\end{equation}
with $N(t)$ the lapse, $a_{i}(t)$ three directional scale factors, and $\omega^{i}$ the left-invariant $\mathfrak{su}(2)$-valued 1-forms on $S^{3}$ satisfying $d\omega^{i}=\tfrac{1}{2}\epsilon^{ijk}\omega^{j}\wedge\omega^{k}$. The self-dual Ashtekar connection aligned with the invariant frame is $\mathcal{A}^{i}=f_{i}(t)\,\omega^{i}$ with $f_{i}=(-\Gamma_{i}+ib_{i})/2$, where the Cartan structure equation $de^{i}+\epsilon^{ijk}\Gamma^{j}\wedge e^{k}=0$ on the diagonal triad $e^{i}=a_{i}\omega^{i}$ fixes the spin connection coefficient $\gamma_{i}=-\Gamma_{i}/2$ with $\Gamma_{i}=a_{j}/a_{k}+a_{k}/a_{j}-a_{i}^{2}/(a_{j}a_{k})$ ($j\neq k\neq i$); $b_{i}(t)$ is the extrinsic curvature part of the connection; since $K^{i}=(\dot a_{i}/N)\,\omega^{i}$ on the diagonal metric \eqref{eq:IX-metric}, the normalization of $f_{i}$ gives $b_{i}=2\dot a_{i}/N$ in Lorentzian evolution. The densitized triad is diagonal with momenta $p_{i}=\tfrac12\,a_{j}a_{k}$ ($j\neq k\neq i$; the factor $\tfrac12$ compensates the $\tfrac12$ in $f_{i}$, so the pair $(b_{i},p_{i})$ is canonical), and the Poisson bracket reads $\{b_{i},p_{j}\}=\delta_{ij}\,\ell_{P}^{2}/V_{c}$ \cite{AHMHH}, with $V_{c}$ the comoving volume of $S^{3}$ in Maurer--Cartan units. The $\mathrm{SU}(2)$ generators are taken Hermitian, with $\mathrm{tr}(T_{i}T_{j})=\delta_{ij}$ and $\mathrm{tr}(T_{i}T_{j}T_{k})=\tfrac{1}{2}\epsilon_{ijk}$.

\textbf{The holomorphic exponent.}---On the Bianchi~IX ansatz, the holomorphic Chern--Simons functional \eqref{eq:YCS} becomes
\begin{equation}
Y_{\rm CS}^{\rm IX}[f]
\;=\;
V_{c}\!\left[\,\sum_{i}\!f_{i}^{2}
\;+\;2\,f_{1}f_{2}f_{3}\,\right] ,
\label{eq:IX-YCSfull}
\end{equation}
which, on the closed-FRW symmetric subspace ($f_{i}=f$), reproduces the sphaleron profile $N_{\rm CS}(f)=3f^{2}+2f^{3}$ \cite{AF}. Substituting $f_{i}=(-\Gamma_{i}+ib_{i})/2$ separates $Y_{\rm CS}^{\rm IX}$ into its real and imaginary parts:
\begin{align}
\mathrm{Re}\,Y_{\rm CS}^{\rm IX}[b]
&\;=\;
\frac{V_{c}}{4}\!\left[\,\sum_{i}(\Gamma_{i}^{2}-b_{i}^{2})
-\Gamma_{1}\Gamma_{2}\Gamma_{3}
+\!\!\!\sum_{(ijk)\,\rm cyc}\!\!\!\Gamma_{i}\,b_{j}b_{k}\,\right],
\label{eq:IX-ReYCS}\\[4pt]
\mathrm{Im}\,Y_{\rm CS}^{\rm IX}[b]
&\;=\;
V_{c}\!\left[\,-\tfrac{1}{2}\!\sum_{i}\!\Gamma_{i}b_{i}
\;+\;\tfrac{1}{4}\!\!\!\!\sum_{(ijk)\,\rm cyc}\!\!\!\!\Gamma_{j}\Gamma_{k}\,b_{i}
\;-\;\tfrac{1}{4}\,b_{1}b_{2}b_{3}\,\right] ,
\label{eq:IX-YCS}
\end{align}
where the sums on $(ijk)$ run over cyclic permutations of $(1,2,3)$. Note that the cubic $b_{1}b_{2}b_{3}$ sits entirely in $\mathrm{Im}\,Y_{\rm CS}^{\rm IX}$ while $\mathrm{Re}\,Y_{\rm CS}^{\rm IX}$ is quadratic in $b$. On the closed-FRW symmetric subspace ($\Gamma_{i}=1$, $b_{i}=b$), \eqref{eq:IX-YCS} collapses to $-\tfrac{V_{c}}{4}(b^{3}+3b)$, where the latter is the standard closed slicing Chern--Simons phase up to an overall sign that is absorbed into the convention of the integration cycle. Note, subleading constants and quantum-torsion contributions \cite{AHMHH} have been suppressed; they shift only the overall normalization and the argument of the Airy function below. Fourier transforming the connection representation Kodama state to the metric ($p$) representation against the kernel $\exp(+iV_{c}p_{i}b_{i}/\ell_{P}^{2})$ (the restriction to the Lorentzian fibre \eqref{eq:fibre} of the canonical kernel of \eqref{eq:Psi-E})\footnote{The explicit $i$ here is not an independent Schr\"odinger phase; it is the canonical kernel $\exp\!\bigl(+\ell_{P}^{-2}\!\int_{\Sigma}\!E\!\cdot\!\mathcal{A}\bigr)$ of \eqref{eq:Psi-E} restricted to the Lorentzian fibre \eqref{eq:fibre}. On the Bianchi~IX ansatz, $\ell_{P}^{-2}\!\int_{\Sigma}E\!\cdot\!\mathcal{A}=(V_{c}/\ell_{P}^{2})\sum_{i}p_{i}(-\Gamma_{i}+ib_{i})$; its $b$-dependent part is $+iV_{c}\,p\!\cdot\!b/\ell_{P}^{2}$, while the $b$-independent real factor $\exp(-V_{c}\sum_{i}p_{i}\Gamma_{i}/\ell_{P}^{2})$ is absorbed into the normalization as in Sec.~\ref{sec:setup}. Thus, the holomorphic (no $i$) transform of Sec.~\ref{sec:setup} becomes an oscillatory Fourier transform once written in the real extrinsic curvature variable $b$.} produces a contour integral whose integrand exponent (\ref{eq:Idef}) becomes
\begin{equation}
I^{\rm IX}[b\,;\,p]
\;=\;
\kappa\,\mathrm{Re}\,Y_{\rm CS}^{\rm IX}(b)
\;+\;
i\,S^{\rm IX}(b;p),
\qquad
S^{\rm IX}\equiv\kappa\,\mathrm{Im}\,Y_{\rm CS}^{\rm IX}(b)+\tfrac{V_{c}}{\ell_{P}^{2}}\,p\!\cdot\!b .
\label{eq:IX-I-full}
\end{equation}
The phase part is conveniently written by absorbing the Fourier source into the same prefactor as $\mathrm{Im}\,Y_{\rm CS}^{\rm IX}$, using
$\kappa=3/(2\ell_{P}^{2}\Lambda)$:
\begin{equation}
i\,S^{\rm IX}[b;p]
\;=\;-\,\frac{i\kappa V_{c}}{4}\,\Bigl[\,
b_{1}b_{2}b_{3}\,+\,2\!\sum_{i}\!\Gamma_{i}b_{i}-\!\!\!\sum_{(ijk)\,\rm cyc}\!\!\!\Gamma_{j}\Gamma_{k}\,b_{i}
\,-\,\tfrac{8\Lambda}{3}\!\sum_{i}\!p_{i}b_{i}\,\Bigr] .
\label{eq:IX-iS}
\end{equation}
Writing $b_{n}=x_{n}+iy_{n}$, which may be thought of as Fresnel rotating the holomorphic variable on the complex configuration space,\footnote{Two points justify this step. First, viewed as a function of the complexified variable $b$, the integrand $e^{I(b)}$ is holomorphic everywhere in the complex $b$-plane: $I(b)$ is a polynomial, so $e^{I(b)}$ has no poles or branch cuts anywhere. With no singularities to cross, Cauchy's theorem lets us rotate the contour $b\to e^{i\theta}r$. Second, this Fresnel rotation is \emph{not redundant}. On the undeformed real axis the trace Hessian is pure imaginary ($\mathrm{Re}\,M_{\rm tr}=0$), so $|e^{I}|=1$ and the integrand merely oscillates (inducing conditional convergence). The rotation by $\theta=\tfrac12(\pi-\arg M)$ is the smallest angle that makes the quadratic $\tfrac12 M(b-b_{*})^{2}$ real and negative, turning oscillation into exponential decay. The shear modes differ: there, $\mathrm{Re}\,M_{\rm sh}=-3\kappa V_{c}/4<0$ already damps the integrand on the real axis, so the rotation is optional and only sharpens the decay rate to $|M_{\rm sh}|$.} allows us to write our Morse function $h^{\rm IX}\equiv\mathrm{Re}\,I^{\rm IX}$ explicitly:
\begin{align}
h^{\rm IX}(x,y)=\frac{\kappa V_{c}}{4}\Bigl[\;
&\underbrace{\textstyle\sum_{i}\bigl(\Gamma_{i}^{2}-x_{i}^{2}+y_{i}^{2}\bigr)
-\Gamma_{1}\Gamma_{2}\Gamma_{3}
+\!\!\sum_{(ijk)\,\rm cyc}\!\Gamma_{i}\bigl(x_{j}x_{k}-y_{j}y_{k}\bigr)}_{\text{modulus}:\;\ \mathrm{Re}\,Y_{\rm CS}^{\rm IX}}
\nonumber\\[2pt]
&+\underbrace{x_{1}x_{2}y_{3}+x_{2}x_{3}y_{1}+x_{3}x_{1}y_{2}-y_{1}y_{2}y_{3}
+\textstyle\sum_{i}\ell_{i}\,y_{i}}_{\text{cubic sector (the phase }S^{\rm IX}\text{ continued off the real axis)}}\Bigr],
\label{eq:IX-Morse-explicit}
\end{align}
with $\ell_{i}\equiv 2\Gamma_{i}-\Gamma_{j}\Gamma_{k}-\tfrac{8\Lambda}{3}p_{i}$ the linear coefficients of \eqref{eq:IX-iS}. The structure is now transparent. On the real $b$-axis ($y=0$), the second group vanishes and $h^{\rm IX}|_{y=0}=\kappa\,\mathrm{Re}\,Y_{\rm CS}^{\rm IX}|_{y=0}$ is quadratic with no cubic, this leads to an oscillatory integrand. The cubic $b_{1}b_{2}b_{3}$ re-enters $h^{\rm IX}$ only off the real axis through $-y_{1}y_{2}y_{3}$ and the mixed terms $x_{i}x_{j}y_{k}$; this is where the Lefschetz thimble lives, and this is why $I^{\rm IX}$ must be continued off the real axis as a whole before its real part is taken since evaluating $\mathrm{Re}\,I^{\rm IX}$ on the real axis alone discards the cubic that produces the Airy direction. We therefore retain the full holomorphic $I^{\rm IX}$ throughout.\footnote{Dropping the modulus $\kappa\,\mathrm{Re}\,Y_{\rm CS}^{\rm IX}$ from \eqref{eq:IX-I-full} gives the pure phase ``real Chern--Simons'' state of \cite{Magueijo,AHMHH}: $\Psi_{\rm K}(b)\propto \exp(iS^{\rm IX})$. That reduction reproduces the de~Sitter saddle and the Airy exactly, but it fails to damp the shear amplitude (Appendix~\ref{app:fullmorse})---the modulus is precisely what supplies the negative real part of the shear Hessian---so we do not use it.}

\textbf{Critical points and branch selection.}---Because $\delta(\mathrm{Re}\,Y_{\rm CS}^{\rm IX})/\delta b$ vanishes on the symmetric configuration (Appendix~\ref{app:fullmorse}), the critical point equation $\partial I^{\rm IX}/\partial b_{i}=0$ reduces to $\partial S^{\rm IX}/\partial b_{i}=0$; this yields the coupled algebraic system
\begin{equation}
b_{j}\,b_{k} \;=\; \tfrac{8\Lambda}{3}\,p_{i}\;+\;\Gamma_{j}\Gamma_{k}\;-\;2\Gamma_{i}, \qquad
i\neq j\neq k .
\label{eq:IX-saddle-eq}
\end{equation}
On closed-FRW symmetric Cauchy data ($a_{1}=a_{2}=a_{3}\equiv a$, $\Gamma_{i}=1$, $p_{i}=p$), the system reduces to $b^{2}=(8\Lambda/3)\,p-1$ which has two branches $b=\pm\sqrt{(8\Lambda/3)p-1}$: the Lefschetz thimble prescription selects the expanding branch
\begin{equation}
b_{1}=b_{2}=b_{3}\equiv b_{*}\;=\;+\sqrt{\tfrac{8\Lambda}{3}p-1},
\qquad
b_{*}^{2}\;=\;\tfrac{8\Lambda}{3}\,p\;-\;1 .
\label{eq:IX-FRW}
\end{equation}
With the dictionary $b_{i}=2\dot a_{i}/N$, $p_{i}=\tfrac12 a_{j}a_{k}$, the saddle relation \eqref{eq:IX-FRW} is precisely the closed Friedmann equation $\dot a^{2}=\tfrac{\Lambda}{3}a^{2}-\tfrac14$ for de~Sitter space in Maurer--Cartan normalization.\footnote{The fiducial $S^{3}$ of \eqref{eq:IX-metric} has spatial curvature $\tfrac14$.} In particular, $b_{*}$ is the rescaled expanding Hubble rate at the de~Sitter saddle, and the $-1$ in \eqref{eq:IX-FRW} comes from the spatial curvature contribution absent in the flat slicing of Sec.~\ref{sec:dS}. Within the selected expanding branch the symmetric saddle is the only critical point for symmetric Cauchy data because the cyclic relation \eqref{eq:IX-saddle-eq} across all three indices forces $b_{1}=b_{2}=b_{3}$ there \footnote{For anisotropic Cauchy data, \eqref{eq:IX-saddle-eq} admits further complex roots, and we work perturbatively about the symmetric saddle}.

\textbf{Saddle point expansion: the Hessian of $\mathrm{Re}\,I$.}---In the semiclassical regime $\kappa V_{c}b_{*}\gg1$, the integral
$Z_{\rm K}^{\rm IX}[p]=\int \exp(I^{\rm IX})d^{3}b$ is dominated by the de~Sitter saddle. Strictly, in Witten's prescription, the Lefschetz thimble is the descending manifold of the de Sitter saddle associated to the Morse function \eqref{eq:IX-Morse-explicit}, so we expand $h^{\rm IX}$ itself to quadratic order about the saddle. In the real coordinates $b_{i}=b_{*}+\delta x_{i}+i\,\delta y_{i}$, $z=(\delta x_{i},\delta y_{i})\in\mathbb{R}^{6}$, the real symmetric Hessian $\mathcal{H}_{ab}=\partial^{2}h^{\rm IX}/\partial z_{a}\partial z_{b}|_{b_*}$ is block diagonal in the eigenbasis $v_{\rm tr}=(1,1,1)/\sqrt3$, $v_{{\rm sh},1}=(1,-1,0)/\sqrt2$, $v_{{\rm sh},2}=(1,1,-2)/\sqrt6$, with real eigenvalues
\begin{equation}
\lambda_{\rm tr}^{\pm}=\pm\frac{\kappa V_{c}b_{*}}{2},
\qquad
\lambda_{\rm sh}^{\pm}=\pm\frac{\kappa V_{c}}{4}\sqrt{9+b_{*}^{2}}\ \ ,
\label{eq:IX-MorseEig}
\end{equation}
respectively. Observe, $\mathcal{H}_{ab}$ is harmonic ($\mathrm{tr}\,\mathcal{H}_{ab}=0$ since $\mathrm{Re}\,I$ is harmonic) of signature $(3,3)$: each mode has one descending eigenvector ($\lambda^{-}<0$, along which $\mathrm{Re}\,I^{\rm IX}$ decreases and $\exp(I^{\rm IX})$ decays) and one ascending eigenvector ($\lambda^{+}>0$). The three descending eigenvectors span the Lefschetz thimble $\mathcal{J}_{*}^{\rm IX}$ and the three ascending ones span the Lefschetz anti-thimble $\mathcal{K}^{\rm IX}_*$.

Two features are decisive. First, the trace curvature of the Morse function along the trace direction $|\lambda_{\rm tr}|=\kappa V_{c}b_{*}/2$ is sourced entirely by the cubic $b_{1}b_{2}b_{3}$ residing in $\mathrm{Im}\,Y_{\rm CS}^{\rm IX}$. Were the modulus $\mathrm{Re}\,Y_{\rm CS}^{\rm IX}$ to act alone, the trace eigenvalue would vanish (Appendix~\ref{app:fullmorse}) and no Airy direction would exist. Second, the eigenvalues identify precisely which direction requires a contour change. On the real-$b$ contour, the shear sector is already convergent since its quadratic coefficient has negative real part,
\begin{equation}
\operatorname{Re} M_{\rm sh}
=
-\frac{3\kappa V_c}{4}
<0 .
\end{equation}
Thus, the shear integrals are Gaussian damped on the original contour and require no change of the integration cycle. The trace sector is different. Its quadratic coefficient is purely imaginary,
\begin{equation}
\operatorname{Re} M_{\rm tr}=0 ,
\end{equation}
so the trace integral is only conditionally convergent on the original contour; this is the direction in which the integration cycle must be continued into the complex plane:
\begin{equation}
b\to x+i y .
\end{equation}
For holomorphic $I^{\rm IX}$ this Morse data is packaged in the complex Hessian
\begin{equation}
M_{ij}\equiv\dfrac{\partial^{2}I^{\rm IX}}{\partial b_{i}\partial b_{j}}\Bigg|_{b_{*}}\!\!\!=\tfrac{\kappa V_{c}}{4}\bigl[(1-ib_{*})\,J_{ij}-(3-ib_{*})\,\delta_{ij}\bigr],
\qquad J_{ij}\equiv1
\label{eq:IX-Hessian-full}
\end{equation}
whose complex eigenvalues
\begin{equation}
M_{\rm tr}=-\tfrac{i\kappa V_{c}b_{*}}{2},
\qquad
M_{\rm sh}=\tfrac{\kappa V_{c}}{4}\bigl(-3+i\,b_{*}\bigr)\ \ (\text{twice})
\label{eq:IX-eigvals}
\end{equation}
have magnitudes equal to the $\mathrm{Re}\,I^{\rm IX}$ curvatures of \eqref{eq:IX-MorseEig} ($|M_{\rm tr}|=|\lambda_{\rm tr}|$, $|M_{\rm sh}|=|\lambda_{\rm sh}|$) and phases that fix the descending eigenvectors below. We use $M_{I}$ purely as a bookkeeping device---it carries no information beyond $\mathrm{Re}\,I$.

\begin{figure}[t]
\centering
\IfFileExists{figure_thimble_3d.pdf}
  {\includegraphics[width=0.99\columnwidth]{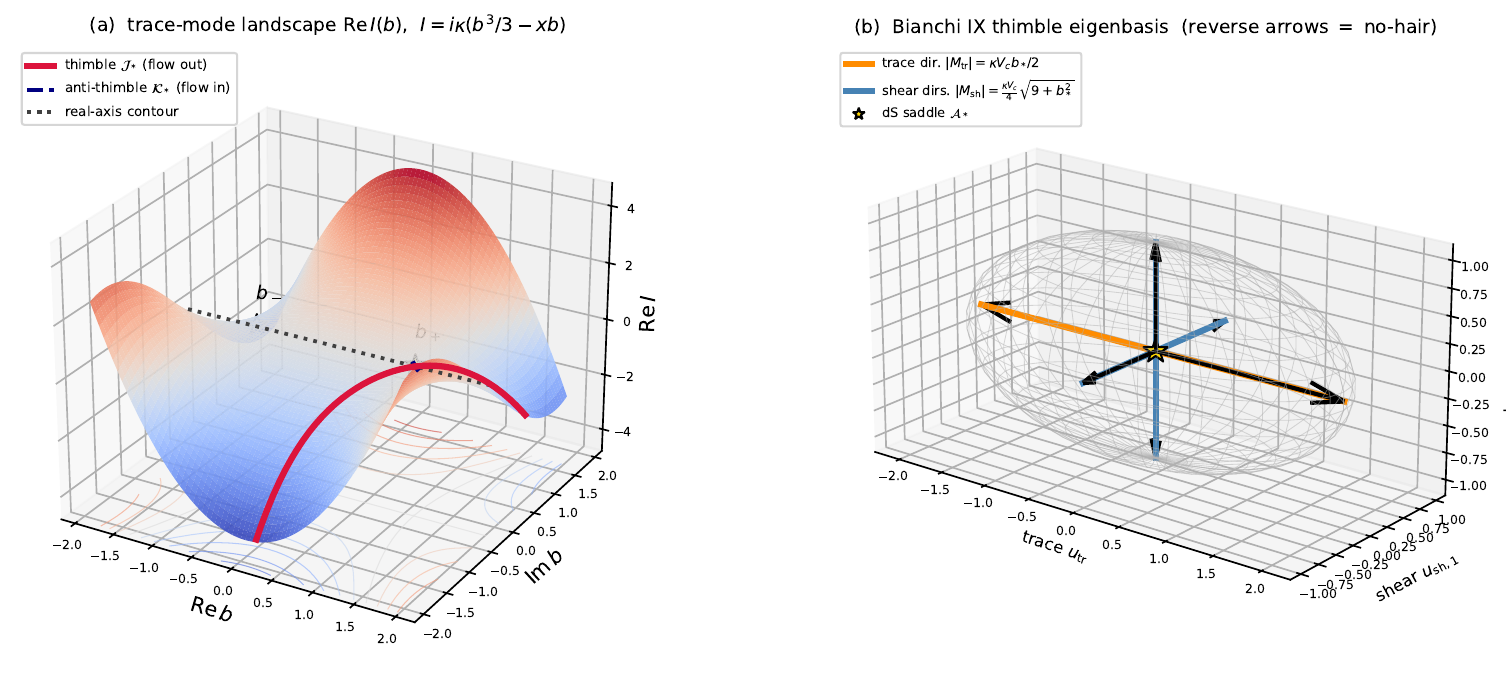}}
  {\fbox{\parbox{0.95\columnwidth}{\centering\vspace{2em}[\,figure\_thimble\_3d.pdf
   --- thimble / gradient-flow figure to be supplied\,]\vspace{2em}}}}
\caption{The Lefschetz thimble through the de~Sitter saddle in the Bianchi~IX truncation. Black arrows show the downwards gradient flow $dz/ds=-\overline{\partial I}$ that sweeps out the Lefschetz thimble. \emph{Left:} the Morse landscape of $\mathrm{Re}\,I$ over the complexified trace direction $b\in\mathbb{C}$ for $I(b)=-i\kappa(b^{3}/3-xb)$ (sign convention as in \eqref{eq:ZKMS}); the gold star is the convergent saddle $b_{+}$ (expanding branch), the white star its conjugate $b_{-}$ (contracting branch). The red curve is the Lefschetz thimble $\mathcal{J}_{*}$ through $b_{+}$; the blue dashed curve is the dual Lefschetz anti-thimble $\mathcal{K}_{*}$; the thin dotted line is the original oscillatory real axis contour. The saddle is thus a hyperbolic fixed point of the flow. \emph{Right:} the same construction in $\mathbb{C}^{3}$, for the Bianchi~IX truncation, plotted in the Hessian eigenbasis (trace $\times$ shear $\times$ shear). The gold star is the convergent saddle $b_+$; the bold orange (trace) and blue (shear) lines are the three Fresnel-rotated thimble eigendirections; and the black arrows show the steepest descent flow emanating from $\mathcal{A}_{*}$ along the thimble. The dashed gray ellipsoid is the bounding initial condition surface elongated 2:1 along the trace direction by the Hessian-eigenvalue ratio of \eqref{eq:IX-eigvals}. Reversing every arrow gives the time-reversed flow under which all anisotropic directions contract onto $\mathcal{A}_{*}$ at rates $|M_{\rm tr}|=\kappa V_{c}b_{*}/2$ (trace) and $|M_{\rm sh}|=\kappa V_{c}\sqrt{9+b_{*}^{2}}/4$ (shear)---the Lefschetz manifestation of cosmic no-hair, in the ratio $2:1$, semiclassically.}
\label{fig:thimble}
\end{figure}

\textbf{Mode-dependent steepest-descent contours.}---The descending eigenvector of mode $I$ is the ray $u_{I}=e^{i\theta_{I}}r_{I}$, $r_{I}\in\mathbb{R}$, along which $\mathrm{Re}\,I^{\rm IX}$ is maximal at the saddle and decreases (equivalently, $M_{I}u_{I}^{2}=-|M_{I}|r_{I}^{2}<0$), which determines the orientation of the steepest-descent ray. We write
\begin{equation}
\theta_{I}=\tfrac12\bigl(\pi-\arg M_{I}\bigr):\qquad
\theta_{\rm tr}=-\tfrac{\pi}{4},\quad
\theta_{\rm sh}=\tfrac12\arctan(b_{*}/3)\to+\tfrac{\pi}{4}\ \ (b_{*}\gg3).
\label{eq:ray}
\end{equation}
The Bianchi IX truncated Lefschetz thimble is the descending manifold of \(\operatorname{Re} I^{\rm IX}\) attached to the chosen saddle; after diagonalizing the fluctuation Hessian into one trace mode and two shear modes, the local form of this Lefschetz thimble is
\begin{equation}
\mathcal{J}_{*}^{\rm IX}
=
\left\{
(u_{\rm tr},u_{{\rm sh},1},u_{{\rm sh},2})\,:\,
u_{\rm tr}\in e^{-i\pi/4}\mathbb{R},
\quad
u_{{\rm sh},\alpha}\in e^{i\theta_{\rm sh}}\mathbb{R},
\quad
\alpha=1,2
\right\}.
\label{eq:IX-thimble}
\end{equation}
Figure~\ref{fig:thimble} shows this cycle using the corresponding upward gradient flow. Moreover, as already mentioned, the trace ray produces the cubic Airy decay while the shear rays produce ordinary Gaussian damping---in particular, the Lefschetz thimble is essential for the trace mode but merely confirms the already converging Gaussian character of the shear modes.

\textbf{Trace mode: Airy.}---Along the descending trace ray of \(\operatorname{Re} I^{\rm IX}\),
\begin{equation}
u_{\rm tr}\in e^{-i\pi/4}\mathbb{R},
\qquad
u_{{\rm sh},\alpha}=0 ,
\end{equation}
the non-exponential modulus is constant (as shown in Appendix~\ref{app:fullmorse}). The falloff of the integrand is therefore controlled entirely by the cubic part of the pure phase action \(iS^{\rm IX}\). Restricting \eqref{eq:IX-I-full} to this one-dimensional ray gives
\begin{equation}
I^{\rm IX}_{\rm tr}
=
I^{\rm IX}_{*}
-
\frac{i\kappa V_c}{4}
\left[
L_{\rm tr} u_{\rm tr}
+
b_{*} u_{\rm tr}^{2}
+
\frac{1}{3\sqrt{3}}u_{\rm tr}^{3}
\right],
\qquad
L_{\rm tr}
=
\sqrt{3}\left(b_{*}^{2}-\bar P\right),
\label{eq:IX-trace-action}
\end{equation}
where
\begin{equation}
\bar P
\equiv
\frac{1}{3}
\sum_i
\left[
\frac{8\Lambda}{3}p_i
+
\Gamma_j\Gamma_k
-
2\Gamma_i
\right].
\label{eq:Pbar-def}
\end{equation}
In the isotropic sector this reduces to
\begin{equation}
\bar P
=
\frac{8\Lambda}{3}p-1 ,
\label{eq:Pbar-symmetric}
\end{equation}
where the \(-1\) comes from the fiducial spatial curvature contribution (cf.\ the paragraph below \eqref{eq:IX-FRW}). Equation~\eqref{eq:IX-FRW} is precisely the saddle condition \(b_{*}^{2}=\bar P\), and hence
\begin{equation}
L_{\rm tr}=0
\end{equation}
at the saddle. The cubic coefficient \(1/(3\sqrt{3})\) is simply the projection of \(b_1b_2b_3\) onto the symmetric trace direction. Thus, the trace reduction of the Bianchi~IX action reproduces the same pure phase cubic structure found in the isotropic model.

The Airy form follows by performing the Fresnel rotation
\begin{equation}
u_{\rm tr}=e^{-i\pi/4}r
\end{equation}
and completing the cube as in Appendix~\ref{app:airy-shift}. Therefore, the trace contribution to the Bianchi~IX Kodama integral becomes
\begin{equation}
Z^{\rm IX}_{K,{\rm tr}}
=
e^{i\Delta_{*}}\,
2\pi\eta'\,
{\rm Ai}\!\left(x_{\rm IX}\right),
\qquad
x_{\rm IX}
=
-
\left(
\frac{3\kappa V_c}{4}
\right)^{2/3}
\bar P .
\label{eq:IX-Airy}
\end{equation}
Here, \(\eta'\) and \(\Delta_{*}\) are defined in Appendix~\ref{app:airy-shift}. The factor \(e^{i\Delta_{*}}\) is the constant stationary phase, fixed by the conservation of \(\operatorname{Im} I^{\rm IX}\) on the Lefschetz thimble, while the Airy function is the convergent integral of \(\exp(\operatorname{Re} I^{\rm IX})\) along the descending trace ray. As in Sec.~\ref{sec:dS}, for $x_{\rm IX}<0$ the real Airy function arises from using both the expanding trace Lefschetz thimble together with its contracting conjugate, both with unit weight; the expanding Lefschetz thimble alone would isolate a single (outgoing) WKB branch.

Several features of \eqref{eq:IX-Airy} are worth emphasizing. First, the \(b_{*}^{2}\) term appearing in \(L_{\rm tr}\) is removed by the cube completion, so the Airy argument depends only on the geometric source \(\bar P\). Second, the exponent \(2/3\) is fixed by dimensional consistency: the source term and the cubic term carry the same overall factor \(\kappa V_c\), so the canonical Airy variable scales as \((\kappa V_c)^{2/3}\bar P\). Finally, the argument vanishes at the turning point
\begin{equation}
\bar P=0
\end{equation}
which corresponds to \(b_{*}=0\) and the minimal-volume configuration. The
expanding de~Sitter saddle has
\begin{equation}
\bar P=b_{*}^{2}>0 ,
\end{equation}
and therefore
\begin{equation}
x_{\rm IX}<0,
\end{equation}
i.e.\ it lies in the oscillatory region of the Airy function, as expected for the Lorentzian de~Sitter branch.

\textbf{Shear modes}---We now evaluate the shear sector contribution by expanding the full exponent \eqref{eq:IX-I-full} to quadratic order about the saddle \eqref{eq:IX-FRW}, along the two shear descending rays of \eqref{eq:IX-thimble},
\begin{equation}
u_{\rm sh}^{(\alpha)}\in e^{i\theta_{\rm sh}}\mathbb{R},
\qquad \alpha=1,2 .
\end{equation}
Write $b_{i}=b_{*}+u_{i}$ with the fluctuation restricted to the traceless (shear) subspace, $\sum_{i}u_{i}=0$. The modulus is quadratic and contributes its Hessian \eqref{eq:ReHess} directly; the only cubic term in \eqref{eq:IX-I-full} is the $b_{1}b_{2}b_{3}$ of \eqref{eq:IX-iS} for which the traceless condition implies $\sum_{i<j}u_{i}u_{j}=-\tfrac12\sum_{i}u_{i}^{2}$ and hence the exact expansion
\begin{equation}
b_{1}b_{2}b_{3}
\;=\;
b_{*}^{3}\;-\;\frac{b_{*}}{2}\sum_{i}u_{i}^{2}\;+\;u_{1}u_{2}u_{3} .
\label{eq:shear-cubic-expansion}
\end{equation}
Three features follow: (1) the quadratic term supplies the $ib_{*}$ part of the shear Hessian $M_{\rm sh}$ in \eqref{eq:IX-eigvals}; (2) no term linear in the shears appears, so the cubic does not displace the saddle; and (3) the only purely shear interaction is the vertex $u_{1}u_{2}u_{3}$. Over the Gaussian width $\sigma_{\rm sh}$ of \eqref{eq:shear-Gaussian} below, this vertex is suppressed, $\kappa V_{c}\,\sigma_{\rm sh}^{3}\sim(\kappa V_{c})^{-1/2}$ in the semiclassical regime $\kappa V_{c}b_{*}\gg1$, so it is an $O(\kappa^{-1/2})$ correction to the Gaussian.\footnote{This is in contrast with the trace cubic, which is retained exactly: the trace Hessian $M_{\rm tr}\propto b_{*}$ degenerates at the turning point $b_{*}=0$, where the two de~Sitter saddles coalesce and the Gaussian approximation fails.} At leading quadratic order, then, each shear mode is quadratic-plus-linear in \eqref{eq:IX-I-full},
\begin{equation}
I^{\rm IX}\big|_{\rm sh}=I^{\rm IX}_{*}
+\sum_{\alpha}\Bigl[\tfrac12 M_{\rm sh}\bigl(u_{\rm sh}^{(\alpha)}\bigr)^{2}
-\tfrac{i\kappa V_{c}}{4}\mathcal{L}_{\rm sh}^{(\alpha)}u_{\rm sh}^{(\alpha)}\Bigr],
\quad
\mathcal{L}_{\rm sh}^{(\alpha)}=-\!\sum_{i}\bigl[\tfrac{8\Lambda}{3}p_{i}+\Gamma_{j}\Gamma_{k}-2\Gamma_{i}\bigr](v_{{\rm sh},\alpha})_{i},
\label{eq:shear-quadratic}
\end{equation}
where the source projection $\mathcal{L}_{\rm sh}^{(\alpha)}$ measures the Cauchy data anisotropy and vanishes in the symmetric sector. Along the ray $u_{\rm sh}^{(\alpha)}=e^{i\theta_{\rm sh}}r$, the quadratic becomes the real Gaussian $-\tfrac{\kappa V_{c}}{8}\sqrt{9+b_{*}^{2}}\,r^{2}$,
\begin{equation}
\int_{\mathbb{R}^{2}}\!d^{2}r\,
e^{-\frac{\kappa V_{c}}{8}\sqrt{9+b_{*}^{2}}\,r^{2}}=\frac{8\pi}{\kappa V_{c}\sqrt{9+b_{*}^{2}}},
\qquad
\sigma_{\rm sh}=\sqrt{\frac{4}{\kappa V_{c}\sqrt{9+b_{*}^{2}}}} ,
\label{eq:shear-Gaussian}
\end{equation}
and completing the square in the linear source term against $M_{\rm sh}$---whose real part is negative---leaves a real damping factor. With $|\mathcal{L}_{\rm sh}|^{2}=\sum_{\alpha}|\mathcal{L}_{\rm sh}^{(\alpha)}|^{2}$, the two shear modes therefore contribute the convergent factor $\exp[-3\kappa V_{c}|\mathcal{L}_{\rm sh}|^{2}/(8(9+b_{*}^{2}))]$.

Combining the trace direction Airy of \eqref{eq:IX-Airy} with the shear Gaussian of \eqref{eq:shear-Gaussian} and absorbing the one-loop fluctuation determinant together with the constant Maslov-plus-cubic-shift phase into the formal normalization $\mathcal{N}$, the leading order wavefunction reduces to
\begin{equation}
\Psi_{\rm K}^{\rm IX}[p]
\;=\;
\mathcal{N}\,\mathrm{Ai}\bigl(x_{\rm IX}\bigr)\,
\exp\!\Bigl[-\,\frac{3\,\kappa V_{c}\,|\mathcal{L}_{\rm sh}|^{2}}{8\,(9+b_{*}^{2})}\Bigr] .
\label{eq:IX-leading}
\end{equation}
In particular, on the Bianchi~IX truncated Lefschetz thimble through the de~Sitter saddle, the leading order Bianchi~IX Chern--Simons wavefunction is the FRW Airy function of a curvature-shifted argument $x_{\rm IX}$ multiplied by a finite real Gaussian in the anisotropy source $\mathcal{L}_{\rm sh}$. In the symmetric sector $|\mathcal{L}_{\rm sh}|^{2}=0$, the anisotropy factor reduces to unity; off-symmetric, it provides genuine real Gaussian damping. At this leading (quadratic, saddle point) order the trace Airy is exact---the modulus has no trace direction contribution---and shear enters only as the convergent Gaussian normalization; the cubic shear self-interaction $u_{1}u_{2}u_{3}$ and the trace-shear couplings are higher-order corrections to \eqref{eq:IX-leading} suppressed by $\kappa^{-1/2}$ in the regime $\kappa V_{c}b_{*}\gg1$.

Now, \eqref{eq:IX-leading} is the central result of this paper. The volume dependence of the Bianchi~IX state is the \emph{same} Airy function as in the isotropic reduction, with the curvature-shifted argument $x_{\rm IX}$---so the Airy form is a property of the Lefschetz thimble itself, not an artifact of the 1D collapse---while anisotropies enter only through a real, convergent Gaussian that suppresses departures from isotropy. The damping factor is the fingerprint of the contour prescription: it exists only because the Lefschetz thimble descends the full $\mathrm{Re}\,I^{\rm IX}$, i.e.\ the pure-phase reduction yields the same Airy but no damping at all (Appendix~\ref{app:fullmorse}).

\textbf{Quantum cosmic no-hair and the Lefschetz thimble.}\label{subsec:nohair}---Classically, the cosmic no-hair theorem states that an initially expanding homogeneous cosmology with $\Lambda>0$, whose matter obeys the dominant and strong energy conditions, approaches de~Sitter space at late times. In particular, the shear and spatial curvature redshift away exponentially, leaving a geometry with ``no hair'' beyond the de~Sitter radius fixed by $\Lambda$~\cite{Wald1983}. The present construction provides a \emph{quantum} version of this statement, one about the wavefunction of the universe rather than about a single Lorentzian trajectory.

The quantum statement is read directly off the leading order Bianchi~IX Kodama wavefunction \eqref{eq:IX-leading}. Beyond the isotropic Airy factor it carries the real Gaussian
\begin{equation}
\bigl|\Psi_{\rm K}^{\rm IX}[p]\bigr|\;\propto\;
\bigl|\mathrm{Ai}(x_{\rm IX})\bigr|\,
\exp\!\Bigl[-\,\tfrac{3\kappa V_{c}\,|\mathcal{L}_{\rm sh}|^{2}}{8\,(9+b_{*}^{2})}\Bigr],
\label{eq:nohair-amplitude}
\end{equation}
so the amplitude assigned to a configuration falls off exponentially with its anisotropy $|\mathcal{L}_{\rm sh}|^{2}$ away from the isotropic (FRW) point where $|\mathcal{L}_{\rm sh}|^{2}=0$. The Chern--Simons--Kodama state is therefore peaked on de~Sitter---anisotropic geometries are present, but exponentially suppressed in amplitude. This is cosmic no-hair realized at the level of the wavefunction, and, in contrast to the global flow statement below, it is rigorous, following from the saddle point evaluation of Sec.~\ref{sec:bianchi} on the Lefschetz thimble through the de~Sitter saddle.

The gradient flow supplies the geometric mechanism behind this suppression.\footnote{Recall, with the downwards gradient flow \eqref{eq:PL-flow} (integrand $e^{I}$, downwards gradient of $\mathrm{Re}\,I$), the Lefschetz thimble is the \emph{descending} manifold of the saddle---its trajectories \emph{emanate} from the saddle---while the anti-thimble is the \emph{ascending} manifold. In particular, the ``attractor'' picture of no-hair refers to the upwards gradient flow.} By definition, $\mathcal{J}_{*}^{\rm IX}$ is swept out by the downwards gradient flow of $\mathrm{Re}\,I^{\rm IX}$:
\begin{equation}
\frac{d z_{I}}{ds}\;=\;-\,\overline{\Bigl(\frac{\partial I^{\rm IX}}{\partial z_{I}}\Bigr)} .
\label{eq:PL-flow}
\end{equation}
Linearizing at the de~Sitter saddle, the Hessian \eqref{eq:IX-Hessian-full} gives $\partial I^{\rm IX}/\partial u_{I}=M_{I}\,u_{I}$ on each eigenmode, so the linearized flow is $du_{I}/ds=-\overline{M_{I}}\,\overline{u_{I}}$. On the Lefschetz thimble \eqref{eq:IX-thimble}, each eigenmode is parametrized by a real coordinate $r_{I}$ through $u_{I}=e^{i\theta_{I}}r_{I}$ with $\theta_{I}=\tfrac12(\pi-\varphi_{I})$ as in \eqref{eq:ray}; writing $M_{I}=|M_{I}|e^{i\varphi_{I}}$, the prefactor collapses, $-\overline{M_{I}}e^{-2i\theta_{I}}=+|M_{I}|$, and the flow becomes the real ODE
\begin{equation}
\frac{d r_{I}}{ds}\;=\;+|M_{I}|\,r_{I} .
\label{eq:IX-flow}
\end{equation}
The downwards gradient flow moves away from the saddle along each eigenmode, so the time-reversed flow contracts every anisotropy mode onto FRW at the Hessian rates
\begin{equation}
|M_{\rm tr}|=\tfrac{\kappa V_{c}b_{*}}{2},
\qquad
|M_{\rm sh}|=\tfrac{\kappa V_{c}}{4}\sqrt{9+b_{*}^{2}}
\;\xrightarrow{\,b_{*}\gg3\,}\;\tfrac{\kappa V_{c}b_{*}}{4}
\label{eq:IX-rates}
\end{equation}
(the trace contracting twice as fast as the shear, semiclassically). The de~Sitter saddle is thus a hyperbolic fixed point of the flow whose unstable manifold is the Lefschetz thimble, and the amplitude suppression \eqref{eq:nohair-amplitude} is precisely the saddle point image of this contraction: the width of the wavefunction in each anisotropy direction is set by $1/|M_{I}|$. In this sense, the quantum no-hair suppression and the Wald-type exponential decay of shear share a single source---the sign of $\mathrm{Re}\,M_{\rm sh}$---now read off from the Hessian rather than the Einstein equations.

A fully global version, such that the reversed flow returns \emph{every} Bianchi~IX configuration to FRW, not merely those in the linearized neighborhood of $b_{+}$, requires the nonlinear flow, and we do not claim it here. Two observations make it plausible. In the symmetric sector with $\Lambda>0$, the source-deformed integrand $I^{\rm IX}$ has no critical points other than $b_{\pm}$ since the cyclic relation \eqref{eq:IX-saddle-eq} forces $b_{1}=b_{2}=b_{3}$ and admits only $b_{*}=\pm\sqrt{(8\Lambda/3)p-1}$; also, the Kasner-type vacuum solutions that drive the classical mixmaster chaos are critical points of the $\Lambda=0$ Hamiltonian constraint, not of $I^{\rm IX}$, so they do not reappear as competing fixed points once $\Lambda>0$. The nonlinear flow thus has no obvious basin into which anisotropic data could be diverted. Establishing that it indeed carries all such data back to $b_{+}$ is a question for the six coupled gradient-flow ODEs, collected in Appendix~\ref{app:numerical}, whose integration we leave to future work.

Finally, the quantum statement should not be conflated with the classical one. Wald's theorem concerns Lorentzian Einstein evolution in cosmological time under energy conditions~\cite{Wald1983}; meanwhile, \eqref{eq:nohair-amplitude} is a statement about the amplitude the Chern--Simons--Kodama state assigns to anisotropic three-geometries. Moreover, $s$ is the steepest descent parameter on $\mathcal{A}_{\mathbb{C}}$---this is related to a Wick-rotated time, but not identical to it. The two statements are compatible, as both single out de~Sitter for $\Lambda>0$, but live in different mathematical structures, and the quantum result here neither derives from, nor requires, the classical one.
\section{Reality conditions}
\label{sec:reality}
The previous section completed the dynamical analysis: we performed the Picard--Lefschetz construction, observed the Airy form is the leading semiclassical wavefunction in both the FRW and Bianchi~IX reductions, and saw the anisotropic data is exponentially driven into the FRW saddle. The remaining objection to the Chern--Simons state, raised in the introduction, is the Lorentzian reality of the complex Ashtekar variable which the standard $L^2$ inner product does not provide. The purpose of this section is to show that this objection is resolved at leading saddle point order by the AHF inner product. Here, the de~Sitter saddle automatically lies on the AHF reality manifold, so the Lefschetz thimble construction and the reality structure pick out the same critical configuration. This is also where the modulus of \eqref{eq:phase-modulus}---retained throughout the dynamics as part of $\mathrm{Re}\,I$---reappears as the intrinsic curvature weight of the norm.

The Lorentzian content of \eqref{eq:ZK} is enforced by the Ashtekar reality conditions \cite{Ashtekar1987},
\begin{equation}
E^{\dagger}=E,
\qquad
\mathcal{A}+\mathcal{A}^{\dagger}=2\,\Gamma(E),
\label{eq:reality}
\end{equation}
through the K\"ahler-style inner product of \cite{AHF} defined as a measure on the doubled cycle $\mathcal{C}\times\bar{\mathcal{C}}\subset\mathcal{A}_{\mathbb{C}}\times\bar{\mathcal{A}}_{\mathbb{C}}$:
\begin{equation}
\langle\Psi|\Psi\rangle_{\rm AHF}
\;=\;\!\int_{\mathcal{C}\times\bar{\mathcal{C}}}\!
\mathcal{D}\mathcal{A}\,\mathcal{D}\bar{\mathcal{A}}\,
e^{-S(\Re\,\mathcal{A})}\,\bigl|\Psi[\mathcal{A}]\bigr|^{2} ,
\label{eq:AHF}
\end{equation}
with $S(\Re\,\mathcal{A})=\!\int e^{i}\wedge de^{i}$ the cotriad Chern--Simons form on the spatial slice and $\bar{\mathcal{C}}$ the complex conjugate of $\mathcal{C}$.

Observe, the de~Sitter saddle is automatically compatible with the AHF reality manifold. Writing $\mathcal{A}_{*}=\Gamma+iK_{\rm dS}$, with $\Gamma$ and $K_{\rm dS}$ Hermitian for real $E$, we see $\mathcal{A}_{*}+\mathcal{A}_{*}^{\dagger}=2\Gamma$ holds identically; moreover, the norm weight $|\Psi_{\rm K}[\mathcal{A}]|^{2}=|\mathcal{N}|^{2}\,\exp(2\kappa\,\mathrm{Re}\,Y_{\rm CS}(\mathcal{A}))$ depends only on $\mathrm{Re}\,Y_{\rm CS}$ (the phase cancels in the modulus); and finally, with $K_{\rm dS}^{i}=He^{i}$, one has $d_{\Gamma}K_{\rm dS}=0$ by torsion-freeness of $\Gamma$, so $\int K_{\rm dS}\wedge D_{\Gamma}K_{\rm dS}=0$ and $\mathrm{Re}\,Y_{\rm CS}(\mathcal{A}_{*})=Y_{\rm CS}(\Gamma)$.

It is the choice of contour that keeps the intrinsic curvature weight in \eqref{eq:AHF} finite, so we state that choice precisely. We take the doubled integration cycle to be the Lefschetz thimble of Secs.~\ref{sec:PL}--\ref{sec:bianchi} through the self-dual de~Sitter saddle and its conjugate, i.e. $\mathcal{C}=\mathcal{J}_{*}$ and $\bar{\mathcal{C}}=\bar{\mathcal{J}}_{*}$. The AHF measure $\exp(-S(\Re\,\mathcal{A}))$ depends only on the real part of the connection, and along $\mathcal{J}_{*}$ that real part is anchored, at the dominant saddle, to $\Re\,\mathcal{A}_{*}=\Gamma(E)$, i.e.\ exactly the AHF reality manifold $\mathcal{A}+\mathcal{A}^{\dagger}=2\Gamma$ of \eqref{eq:reality}. Because the Lefschetz thimble is pinned there by downwards gradient flow, the directions along which $\Re\,\mathcal{A}$ can move off $\Gamma$ are exactly those of Sec.~\ref{sec:bianchi}: the two anisotropic shear directions which the Lefschetz thimble real-Gaussian damps, and the isotropic trace direction controlled by the convergent Airy. It is the shear Gaussian damping---supplied, consistently, by the same modulus $\mathrm{Re}\,Y_{\rm CS}$ that weights this norm---that prevents $\Re\,\mathcal{A}$ from running off to the large, highly anisotropic or degenerate triads on which the intrinsic curvature weight $Y_{\rm CS}(\Gamma)$ would diverge, while the trace direction holds the slice at the finite de~Sitter volume.

The leading saddle point evaluation of the AHF norm is then
\begin{equation}
\langle\Psi_{\rm K}|\Psi_{\rm K}\rangle_{\rm AHF}^{\,\rm leading}
\;=\;|\mathcal{N}|^{2}\,
\exp\!\Bigl[\,2\kappa\,Y_{\rm CS}(\Gamma)
\;-\;\!\int_{\Sigma}\!e^{i}\wedge de^{i}\,\Bigr] ;
\label{eq:AHF-saddle}
\end{equation}
this is real, positive, and in fact finite because both intrinsic geometry terms in the exponent are finite on our compact Cauchy hypersurface $\Sigma=S^3$. Concretely, for flat-FRW, $\Gamma=0$ and $de^{i}=0$, so \eqref{eq:AHF-saddle} reduces to $|\mathcal{N}|^{2}$, normalizing the Chern--Simons state to unit AHF norm at leading order; for the closed Bianchi IX slicing $\Sigma=S^{3}$ with the unit Maurer--Cartan cotriad ($e^{i}=\omega^{i}$), $\int_{S^{3}}e^{i}\wedge de^{i}=3V_{S^{3}}$ (in the orientation conventions of Sec.~\ref{sec:bianchi}; for $e^{i}=a\,\omega^{i}$ the term scales as $a^{2}$) is finite and $Y_{\rm CS}(\Gamma_{S^{3}})$ is the (finite) Chern--Simons invariant of the round $S^3$. Hence, the intrinsic curvature weight is a finite number fixed by the slice geometry rather than a divergence. Thus, the Lefschetz thimble $\mathcal{J}_{*}$ and the AHF reality manifold pick out the same critical configuration in $\mathcal{A}_{\mathbb{C}}$, i.e. the contour that renders the intrinsic curvature contribution finite is precisely the steepest descent thimble through the de~Sitter saddle, and on it the AHF norm of the Chern--Simons state is finite and positive at leading semiclassical order.

This leading-order statement dovetails with the nonperturbative normalizability analysis of Bernardo, Kuntzleman, Pezzelle et al. \cite{ABKP}. Starting from the holomorphic inner product derived from the same reality conditions \eqref{eq:reality}, it was shown there that the graviton linearization of the Chern--Simons--Kodama state is perturbatively normalizable for super-Planckian cosmological constant, and that a phase-space rotation generalizing Thiemann's complexifier renders the full perturbative state normalizable for all $\Lambda$. The saddle point finiteness of \eqref{eq:AHF-saddle} is the minisuperspace counterpart of that statement, and the Lefschetz thimble prescription used here supplies the integration cycle on which the two computations can be compared mode by mode.
\section{Discussion}
\label{sec:disc}
The construction is now in place: the contour selection, the anisotropic Bianchi~IX evidence that the Airy form survives anisotropies, and the AHF reality structure at leading order. The purpose of this section is to summarize the central conceptual result, situate it among related approaches in the literature, and identify the open questions that remain.

The main point is that the Chern--Simons state should not be regarded as an intrinsically ill-defined wavefunctional. Rather, it is a holomorphic Chern--Simons generating functional whose definition depends crucially on the choice of integration cycle. On the real contour, it is defined at best conditionally and appears non-normalizable; on the Lefschetz thimble through the de~Sitter saddle, defined by downwards gradient flow of Witten's Morse function, it is absolutely convergent and admits a controlled large-$\kappa$ expansion. The apparent pathology is therefore not a failure of the state itself, but a failure to specify an allowed integration cycle (which is a part of the definition of the state). In this sense, the role of the Lefschetz thimble is directly analogous to the role of $i\epsilon$ prescriptions and steepest descent contours in ordinary quantum field theory.

The new result of this paper is that the Airy form of the wavefunction is not a 1D artifact. In both reductions analyzed here, FRW and Bianchi~IX, it is the dominant structure on the Lefschetz thimble---anisotropic deviations are exponentially damped by the Lefschetz manifestation of cosmic no-hair derived in Sec.~\ref{subsec:nohair}. This damping is a \emph{real} Gaussian precisely because the contour is defined by descent of the full $\mathrm{Re}\,I$, rather than by the oscillatory phase alone. In particular, the real part of the exponent, $\kappa\,\mathrm{Re}\,Y_{\rm CS}$, supplies the negative real part of the shear Hessian. Thus, the same analytic continuation that renders the Chern--Simons state normalizable also selects the isotropizing branch of the anisotropic theory.

This result also clarifies the relation between the present construction and the isotropization mechanism proposed in \cite{AlexanderAlexandreDanielssonSpergel}. In that work, Bianchi~IX plays the role of a gravitational sphaleron, i.e. an unstable configuration separating anisotropic sectors whose quantum dynamics drives the system toward the isotropic de~Sitter configuration. The present Picard--Lefschetz analysis gives a complementary contour-theoretic realization of the same physical idea. The Bianchi~IX saddle is not merely a minisuperspace curiosity, it is the local representative of the nontrivial configuration space structure through which anisotropic perturbations are funneled into the isotropic branch. The negative directions associated with shear are treated by the Lefschetz thimble rotation while the stable isotropic direction retains the Airy structure. In this language, the sphaleron mechanism appears as a statement about the topology of the complexified integration cycle since anisotropic configurations lie along directions of steepest descent away from the Bianchi~IX saddle and the gradient flow suppresses them relative to the isotropic de~Sitter endpoint. The ``quantum no-hair'' behavior found here is therefore the Lefschetz thimble version of the sphaleron-induced isotropization mechanism. Indeed, the agreement is quantitative: restricting the Plebanski contour prescription of \cite{AlexanderAlexandreDanielssonSpergel} to Bianchi~IX and isolating the boundary Chern--Simons--Kodama functional as in the Appendix of that work, the steepest descent contours chosen there agree with the Lefschetz thimbles constructed here, now with the underlying Picard--Lefschetz theory defined explicitly.

This places the present construction in a definite relation to recent work of \cite{AHMHH} which computed the Bianchi~I and Bianchi~IX Chern--Simons states, by Fourier transform on the real $b_{i}$ axis, and obtained Bessel-function wavefunctions of the symmetric volume direction. The real axis contour passes through both saddles and decomposes, in Picard--Lefschetz language, as the integer-weighted sum of the expanding branch Lefschetz thimble and its contracting branch conjugate. The Bessel form arises from combining the two cubic Airy contributions through the standard $\{\mathrm{Ai},\mathrm{Bi}\}\leftrightarrow\{I_{\pm1/3},K_{1/3}\}$ connection formulas (DLMF~\S 9.6). The prescription used here evaluates the same decomposition on the full integrand $e^{I}$: in the classically forbidden regime the real contour is changed to the single expanding branch Lefschetz thimble, while in the oscillatory regime both conjugate Lefschetz thimbles carry unit weight and combine into the real Airy structure of \eqref{eq:IX-leading}. This is the convergent steepest descent definition prescribed by Witten's analytic continuation of Chern--Simons theory at general level \cite{Witten2010}, and the expanding branch Lefschetz thimble by itself isolates a single physically relevant cosmological branch.

Two conceptual points sharpen this picture. The first concerns large gauge invariance. As discussed in Sec.~\ref{sec:PL}, lifting to the universal cover of the gauge orbit space makes the Lefschetz thimble integrals well-defined, but, in the Lorentzian theory, large gauge transformations, the state is rescaled by a real factor under  so norms built from $|\Psi_{\rm K}|^{2}$ are not by themselves large-gauge invariant. The natural resolution is the embedding of \cite{AlexanderAlexandreDanielssonSpergel}: the Chern--Simons--Kodama functional is the boundary functional of a gauge-invariant state of the Plebanski path integral, with the gauge variance of the boundary compensated by the bulk in the manner of anomaly inflow. The second point is that this embedding, taken together with the results obtained here, supports the identification of the Chern--Simons--Kodama functional as the boundary functional of a generalized, complexified Hartle--Hawking state in the sense of \cite{AHMHH}. Three independent computations now align: the FRW reduction reproduces the Hartle--Hawking Airy wavefunction on the doubled Lefschetz thimble; the real-contour Bianchi wavefunctions of \cite{AHMHH} are recovered as specific members of the Lefschetz thimble decomposition \eqref{eq:thimble-decomp}; and the Bianchi~IX steepest descent contours coincide with those chosen in the Appendix of \cite{AlexanderAlexandreDanielssonSpergel}. The full status of this identification rests on the modes beyond Bianchi~IX, but the boundary functional embedding already indicates that the correspondence should persist there, and we expect a future rigorous analysis to bear this out.

Several open questions persist. The most pressing is the multimode Picard--Lefschetz problem in the full inhomogeneous theory, i.e. identifying the relevant Lefschetz thimbles. In particular, the perturbative inhomogeneous modes about the de~Sitter saddle---the sector in which the negative-norm and negative-energy concerns of \cite{Witten2003} arise---are not addressed by the minisuperspace analysis of this paper; see \cite{ABKP} for a perturbative normalizability analysis of precisely this graviton sector. Extending the Lefschetz thimble construction to that sector is the critical open test. Closely related problems include: the explicit one-loop determinant in the de~Sitter background, the gravitational analog of Witten's Reidemeister torsion and $\eta$-invariant formula \cite{Witten2010}, the BRST gauge fixing of $\mathrm{SL}(2,\mathbb{C})$ in the presence of the source, and the matching of the construction to the Ashtekar reality conditions on physical observables. A further problem in this list is the nonperturbative, gauge-invariant completion of the boundary functional within the Plebanski path integral of \cite{AlexanderAlexandreDanielssonSpergel}; this would make the anomaly inflow resolution of the large gauge dependence fully explicit.

Two natural symmetry-reduced extensions also suggest themselves. The first is Bianchi~I with quantum torsion, where the authors \cite{AHMHH} obtain a closed-form Bessel result that should match the corresponding sum over Lefschetz thimbles. The second is Kantowski--Sachs; here, the symmetry structure differs and the saddle analysis must be redone. In each case, the analytically-continued framework converts the problem into a well-posed multidimensional Picard--Lefschetz problem. The broader implication is that the Kodama/Chern--Simons state, once equipped with its correct Lefschetz thimble prescription and reality structure, provides not only a candidate semiclassical wavefunction of de~Sitter space, but also a mechanism by which anisotropic gravitational configurations are dynamically and quantum mechanically driven toward the isotropic universe.
\acknowledgments
The authors especially thank S. James Gates for inspiring us to look at the mathematical connection between the Complefixied Chern Simons theory and the Kodama state. We also thank Bruno Alexandre, Daine Danielson, Heliudson Bernardo, Keshav Dasgupta, Laurent Freidel, Antal Jevicki, Jacob Kuntzleman, Peiran Liu, Joao Magueijo, Sav Sethi and David Spergel for discussions.  S.A was
supported by The Simons Foundation Target Grant.  K.B was partially supported by an NSF Graduate Research Fellowship award during this work.
\appendix
\section{Bianchi IX integrand and Airy reduction}
\label{app:airy-shift}
This appendix records the two technical steps used in Sec.~\ref{sec:bianchi}: (i)~the derivation of the integrand of the contour integral from the Chern--Simons functional, and (ii)~the explicit cubic-shift change of variables that brings the trace-mode integral to canonical Airy form \eqref{eq:IX-Airy}.

\textit{(i) Contour integral integrand.}---On the Bianchi~IX ansatz, $\mathcal{A}^{i}=f_{i}\omega^{i}$ with $f_{i}=(-\Gamma_{i}+ib_{i})/2$. Evaluating the holomorphic Chern--Simons functional \eqref{eq:YCS} on a constant-$t$ Cauchy slice and separating real and imaginary parts gives \eqref{eq:IX-ReYCS}--\eqref{eq:IX-YCS}. The full Kodama wavefunction in the connection representation is $\Psi_{\rm K}(b)=\mathcal{N}\,\exp(\kappa Y_{\rm CS}^{\rm IX}(b))$, and its metric representation Fourier dual against the canonical kernel $\exp(+iV_{c}\,p_{i}b_{i}/\ell_{P}^{2})$ produces the integrand $\exp(I^{\rm IX}[b;p])$ of \eqref{eq:IX-I-full}.

\emph{Origin of the $8\Lambda/3$ source coefficient.}---The factor is fixed by the: AHMHH bracket, kernel, and coupling $\kappa$. The diagonal AHMHH Poisson bracket $\{b_{i},p_{j}\}=\delta_{ij}\ell_{P}^{2}/V_{c}$ (Sec.~\ref{sec:bianchi}, after \cite{AHMHH}) makes the canonical connection-to-triad kernel $\exp(+iV_{c}p_{i}b_{i}/\ell_{P}^{2})$. (Note, the prefactor $V_{c}/\ell_{P}^{2}$ is exactly what renders $V_{c}p_{i}b_{i}/\ell_{P}^{2}$ the dimensionless generator conjugate to $b_{i}$ under that bracket.) The Chern--Simons phase carries the prefactor $\kappa V_{c}/4$ (Eq.~\eqref{eq:IX-iS}), so relative to it the source coefficient is
\begin{equation}
\frac{V_{c}/\ell_{P}^{2}}{\kappa V_{c}/4}
\;=\;\frac{4}{\kappa\,\ell_{P}^{2}}
\;=\;\frac{4}{3/(2\Lambda)}
\;=\;\frac{8\Lambda}{3},
\label{eq:app-8Lover3}
\end{equation}
using $\kappa\ell_{P}^{2}=3/(2\Lambda)$. This is how the source enters \eqref{eq:IX-iS} with the same $-\tfrac{i\kappa V_{c}}{4}$ prefactor as $\mathrm{Im}\,Y_{\rm CS}^{\rm IX}$, multiplying $\tfrac{8\Lambda}{3}\sum_{i}p_{i}b_{i}$. This is also consistent with the full theory source $J^{ai}=E^{ai}/\ell_{P}^{2}$ of \eqref{eq:JE}: restricted to the Lorentzian fibre \eqref{eq:fibre}, $\ell_{P}^{-2}\!\int_{\Sigma}E\!\cdot\!\mathcal{A}$ reduces to $(V_{c}/\ell_{P}^{2})\sum_{i}p_{i}(-\Gamma_{i}+ib_{i})$, with $p_{i}=\tfrac12 a_{j}a_{k}$, whose $b$-dependent part is exactly the kernel above.

\textit{(ii) Cubic-shift reduction to Airy form.}---Restricting \eqref{eq:IX-I-full} to the trace direction $b_{i}=b_{*}+u_{\rm tr}/\sqrt{3}$ gives, since the modulus is constant there, the single variable cubic \eqref{eq:IX-trace-action} which we write (matching the main-text prefactor exactly)
\begin{equation}
I^{\rm IX}\!\bigl|_{\rm tr}
\;=\;I_{*}^{\rm IX}\;-\;\frac{i\kappa V_{c}}{4}\bigl[\mathcal{L}_{\rm tr}\,u_{\rm tr}
\;+\;b_{*}\,u_{\rm tr}^{2}\;+\;c_{3}\,u_{\rm tr}^{3}\bigr],
\qquad c_{3}=\tfrac{1}{3\sqrt3} .
\label{eq:app-traceprefac}
\end{equation}
Complete the cube directly. With cubic, quadratic, and linear coefficients $\alpha=-\tfrac{i\kappa V_{c}}{4}c_{3}$, $\beta=-\tfrac{i\kappa V_{c}}{4}b_{*}$, $\gamma=-\tfrac{i\kappa V_{c}}{4}\mathcal{L}_{\rm tr}$, the shift
\begin{equation}
u_{\rm tr}\;=\;w\;-\;\frac{\beta}{3\alpha}\;=\;w\;-\;\sqrt{3}\,b_{*}
\label{eq:app-shift}
\end{equation}
removes the quadratic and sends the bracket of \eqref{eq:app-traceprefac} to $c_{3}w^{3}-\sqrt3\,\bar P\,w+\text{const}$. (Note, the $b_{*}^{2}$ of $\mathcal{L}_{\rm tr}=\sqrt3(b_{*}^{2}-\bar P)$ cancels against $-b_{*}^{2}/3c_{3}$.) The exponent is thus $\alpha w^{3}+\gamma' w+\text{const}$ with shifted linear coefficient
\begin{equation}
\gamma'\;=\;\gamma-\frac{\beta^{2}}{3\alpha}\;=\;i\,\tfrac{\kappa V_{c}}{4}\sqrt3\,\bar P .
\label{eq:app-gammaprime}
\end{equation}
Rescaling $w=\eta' t$ to canonicalize the cubic, $\alpha(\eta')^{3}=i/3$, gives
\begin{equation}
(\eta')^{3}\;=\;\frac{i}{3\alpha}\;=\;-\,\frac{4\sqrt3}{\kappa V_{c}},
\qquad
|\eta'|^{3}=\frac{4\sqrt3}{\kappa V_{c}}.
\label{eq:app-etaprime}
\end{equation}
This is a complex cube root whose branch carries the contour onto the Lefschetz thimble (the Fresnel ray $u_{\rm tr}\in e^{-i\pi/4}\mathbb{R}$ of \eqref{eq:ray} near the saddle), after which the trace integral takes the canonical form $Z_{\rm K}^{\rm IX}|_{\rm tr}=e^{i\Delta_{*}}\,2\pi\eta'\,\mathrm{Ai}(x_{\rm IX})$ with $\Delta_{*}$ the constant action shift generated by \eqref{eq:app-shift} and Airy argument
\begin{equation}
x_{\rm IX}\;=\;\frac{\gamma'\,\eta'}{i}
\;=\;-\,\bigl(\tfrac{3\kappa V_{c}}{4}\bigr)^{2/3}\bar P ,
\label{eq:app-xIX}
\end{equation}
which is real on the thimble (the branch of $\eta'$ is fixed by reality of $x_{\rm IX}$) and vanishing at the turning point $\bar P=0$. This is \eqref{eq:IX-Airy} of the main text.
\section{Picard--Lefschetz gradient flow and the Kapustin--Witten equation}
\label{app:gradflow}
This appendix derives Eq.~\eqref{eq:PL-grad} and shows that, on $\Sigma\times\mathbb{R}_{+}$, in temporal gauge, it becomes the Kapustin--Witten equation \eqref{eq:KW}.

\textit{(i) Gradient flow on a K\"ahler manifold.}---Let $X$ be a complex manifold with K\"ahler metric $ds^{2}=g_{a\bar b}\,dz^{a}\,d\bar z^{\bar b}$ and
$f:X\to\mathbb{R}$ a smooth function. The steepest descent flow of $f$ is
\begin{equation}
\frac{dz^{a}}{ds}\;=\;-\,g^{a\bar b}\,\partial_{\bar b}f .
\label{eq:app-grad-flow-Kahler}
\end{equation}
\textit{(ii) Holomorphic functional.}---For $f=\mathrm{Re}\,I$ with $I$ holomorphic, $\partial_{\bar a}I=0$, so $\partial_{\bar b}(\mathrm{Re}\,I)=\tfrac{1}{2}\overline{\partial_{b}I}$. Substituting into \eqref{eq:app-grad-flow-Kahler} and absorbing the factor $\tfrac{1}{2}$ into $s\to 2s$ gives $dz^{a}/ds=-g^{a\bar b}\,\overline{\partial_{b}I}$, which, for a flat metric, is $\dot z^{a}=-\overline{\partial_{a}I}$. This makes the second equality of \eqref{eq:PL-grad} explicit.

\textit{(iii) Field-theory version.}---Promoting $z^{a}\to\mathcal{A}_{i}(x)$
with the flat $L^{2}$ K\"ahler metric
\begin{equation}
\langle\delta\mathcal{A},\delta\mathcal{A}'\rangle
=\int_{\Sigma}\mathrm{tr}(\delta\mathcal{A}_{i}\,\overline{\delta\mathcal{A}'^{\,i}})
\end{equation}
gives
\begin{equation}
\frac{d\mathcal{A}_{i}(x)}{ds}\;=\;-\,\overline{\frac{\delta I[\mathcal{A};J]}{\delta\mathcal{A}_{i}(x)}} ;
\label{eq:app-grad-field}
\end{equation}
this is \eqref{eq:PL-grad}.

\textit{(iv) Reduction to the curvature.}---Variation of $Y_{\rm CS}$ gives
$\delta Y_{\rm CS}=2\int_{\Sigma}\mathrm{tr}(\delta\mathcal{A}\wedge F[\mathcal{A}])$, so $\delta Y_{\rm CS}/\delta\mathcal{A}_{i}=\epsilon^{ijk}F_{jk}[\mathcal{A}]$ and $\delta I/\delta\mathcal{A}_{i}=\kappa\,\epsilon^{ijk}F_{jk}+J^{i}$. The downwards gradient flow
\eqref{eq:app-grad-field} reads
\begin{equation}
\frac{d\mathcal{A}_{i}}{ds}
\;=\;-\,\kappa\,\epsilon^{ijk}\,\overline{F_{jk}[\mathcal{A}]}\;-\;\overline{J^{i}} .
\label{eq:app-flow-explicit}
\end{equation}
\textit{(v) Splitting into Kapustin--Witten form.}---Write $\mathcal{A}=A+i\phi$ with $A,\phi$ real $\mathfrak{su}(2)$-valued. Also, $\overline{F[\mathcal{A}]}=F[A]-\phi\wedge\phi-i\,d_{A}\phi$. Real and imaginary parts of \eqref{eq:app-flow-explicit} at $J=0$ (with the transversality condition $d_{A}\!\star\!\phi=0$) are, viewed as PDEs on $\Sigma\times\mathbb{R}_{+}$ in temporal gauge, the Kapustin--Witten equations \cite{KapustinWitten,Witten1101} at the parameter $\tau$; for $J\neq0$ the constant term $-\bar J$ appears as an inhomogeneous deformation, cf.\ the paragraph below\eqref{eq:KW}. Under the Ashtekar$\,=\,$Witten dictionary \eqref{eq:dictionary} ($A=\Gamma$, $\phi=K$) they reduce to \eqref{eq:KW}. Thus, the Lefschetz thimble $\mathcal{J}_{\alpha}$ through the saddle $\mathcal{A}_{\alpha}$ is the moduli space of Kapustin--Witten solutions on $\Sigma\times\mathbb{R}_{+}$ with $\mathcal{A}\to\mathcal{A}_{\alpha}$ as $s\to-\infty$.
\section{The Morse function $\mathrm{Re}\,I$ and the role of $\mathrm{Im}\,Y_{\rm CS}$}
\label{app:phasechoice}
This appendix records the algebraic identity relating the Morse function $\mathrm{Re}\,I$ to $\mathrm{Im}\,Y_{\rm CS}$, and the bookkeeping that places the modulus in the reality norm. None of the dynamics uses anything beyond $\mathrm{Re}\,I$; the identity simply explains why $\mathrm{Im}\,Y_{\rm CS}$ appears.
On the Lorentzian fibre \eqref{eq:fibre}, the source coupling is purely imaginary; write $I=\kappa\,\mathrm{Re}\,Y_{\rm CS}+iS$ with $S=\kappa\,\mathrm{Im}\,Y_{\rm CS}+(\text{Fourier phase})$ real. Then, on the fibre, $\mathrm{Re}\,I=\kappa\,\mathrm{Re}\,Y_{\rm CS}$ (quadratic, no cubic) and $\mathrm{Im}\,I=S$. If one further drops the modulus---the ``real-Chern--Simons'' state of \cite{Magueijo,AHMHH}---the holomorphic exponent becomes $\tilde I=iS$ and
\begin{equation}
\mathrm{Re}(\tilde I)=\mathrm{Re}(iS)=-\,\mathrm{Im}\,S,\qquad
\mathrm{Im}(\tilde I)=\mathrm{Re}\,S.
\label{eq:pc-identity}
\end{equation}
In particular, working with $\mathrm{Im}\,Y_{\rm CS}$ in the pure phase reduction is \emph{identical} to using a Morse function $\mathrm{Re}(\tilde I)$. Off the fibre, both $\mathrm{Re}\,I$ (full) and $\mathrm{Re}(\tilde I)$ (reduced) are nontrivial height functions carrying the cubic; they coincide on the de~Sitter saddle and the trace direction, but they differ in the shear sector (Appendix~\ref{app:fullmorse}). The main text uses the full $\mathrm{Re}\,I$.

\textit{(ii) Where the modulus goes.}---The modulus $\exp(\kappa\,\mathrm{Re}\,Y_{\rm CS})$ is not discarded; it is (a) retained in the dynamics as the real part of $I$---this is what gives the shear Hessian its negative real part (Eq.~\eqref{eq:IX-eigvals})---and (b) carried into the AHF norm where, at the saddle, $d_{\Gamma}K_{\rm dS}=0$ gives $\mathrm{Re}\,Y_{\rm CS}(\mathcal{A}_{*})=Y_{\rm CS}(\Gamma)$, i.e. the intrinsic curvature weight of \eqref{eq:AHF-saddle}. The $\Gamma$-dependent linear piece of $\mathrm{Im}\,Y_{\rm CS}$ (e.g.\ the $+3b$ of $S^{3}$) stays in $S$ and shifts the Airy argument through $\bar P$ in \eqref{eq:IX-Airy}.
\section{Full vs.\ pure phase Morse function: trace identity, shear difference}
\label{app:fullmorse}
The main text descends the full Morse function $\mathrm{Re}\,I^{\rm IX}$ with $I^{\rm IX}=\kappa Y_{\rm CS}^{\rm IX}+(\text{source})$. The pure phase reduction $\tilde I^{\rm IX}=iS^{\rm IX}$ drops the modulus $\kappa\,\mathrm{Re}\,Y_{\rm CS}^{\rm IX}$. This appendix shows the two share the saddle and trace-mode Airy exactly, but differ in the shear sector. From \eqref{eq:IX-ReYCS}, the modulus is purely quadratic in $b$ (the cubic lies purely in the imaginary part) with gradient and Hessian at the symmetric saddle given by
\begin{equation}
\partial_{b_{i}}\mathrm{Re}\,Y_{\rm CS}^{\rm IX}\big|_{\rm sym}
=\tfrac{V_{c}}{4}(-2b_{*}+2b_{*})=0,
\qquad
\bigl(\partial^{2}\mathrm{Re}\,Y_{\rm CS}^{\rm IX}\bigr)_{ij}
=\tfrac{V_{c}}{4}(1-3\delta_{ij}) ;
\label{eq:ReHess}
\end{equation}
the Hessian has eigenvalues $\{0_{\rm tr},-\tfrac{3V_{c}}{4},-\tfrac{3V_{c}}{4}\}$.

\textit{Saddle unchanged.}---The modulus gradient vanishes at the symmetric configuration, so the de~Sitter critical point $b_{*}$ of \eqref{eq:IX-saddle-eq} is a critical point of the full $I^{\rm IX}$.

\textit{Trace mode unchanged.}---The modulus has vanishing trace direction eigenvalue and is constant on the FRW line ($\mathrm{Re}\,Y_{\rm CS}^{\rm IX}=V_{c}/2$ there), contributing no slope, no curvature, and no cubic along $u_{\rm tr}$. Hence, the trace reduction \eqref{eq:IX-trace-action} and the Airy $\mathrm{Ai}(x_{\rm IX})$ are identical in the two prescriptions.

\textit{Shear modes: the modulus is decisive.}---The full shear Hessian is $M_{\rm sh}=\tfrac{i\kappa V_{c}b_{*}}{4}+\kappa(-\tfrac{3V_{c}}{4})=\tfrac{\kappa V_{c}}{4}(-3+ib_{*})$, Eq.~\eqref{eq:IX-eigvals}. Its negative real part $-3\kappa V_{c}/4$ is a genuine, $b_{*}$-independent Gaussian damping. In the pure phase reduction, the modulus is absent and completing the square against it gives the shear-source factor a purely imaginary exponent---a phase of unit modulus with no amplitude suppression. Thus, the real anisotropy damping of \eqref{eq:IX-leading} exists \emph{only} in the full $\mathrm{Re}\,I$ prescription. The two prescriptions agree on the configuration space Gaussian width as $b_{*}\to\infty$ ($\sqrt{9+b_{*}^{2}}\to b_{*}$), but the pure phase does not reproduce the real source factor; this is the reason the main text retains the full modulus.
\section{Full Bianchi~IX gradient flow: numerical thimble (work in progress)}
\label{app:numerical}
\emph{This appendix records the equations underlying an ongoing numerical study; the integration and its convergence checks are not yet complete, and the results below are stated as a program rather than as established theorems.} Allowing the three diagonal connection components to be independently complex, $b_{i}=x_{i}+i\,y_{i}$, the phase part of the integrand is $iS^{\rm IX}=-\tfrac{i\kappa V_{c}}{4}P(b)$ with
\begin{equation}
P(b)=b_{1}b_{2}b_{3}+2\sum_{i}\Gamma_{i}b_{i}-\!\sum_{i}\Gamma_{j}\Gamma_{k}b_{i}
-\tfrac{8\Lambda}{3}\sum_{i}p_{i}b_{i}.
\label{eq:app-num-P}
\end{equation}
For the numerical experiment, we integrate the full flow $db_{i}/ds=-\overline{\partial I^{\rm IX}/\partial b_{i}}$. The holomorphic derivative splits into a phase piece and a modulus piece,
\begin{equation}
\frac{\partial I^{\rm IX}}{\partial b_{i}}
=\underbrace{-\frac{i\kappa V_{c}}{4}\bigl(b_{j}b_{k}+c_{i}\bigr)}_{\text{phase}}
\;+\;\underbrace{\frac{\kappa V_{c}}{4}\bigl(-2b_{i}+\Gamma_{j}b_{k}+\Gamma_{k}b_{j}\bigr)}_{\text{modulus}},
\qquad
c_{i}\equiv 2\Gamma_{i}-\Gamma_{j}\Gamma_{k}-\tfrac{8\Lambda}{3}p_{i},
\label{eq:app-num-complex-flow}
\end{equation}
with $c_{i}$ real and $(i,j,k)$ the three distinct indices. The modulus piece vanishes at the symmetric saddle ($b_{i}=b_{*}$, $\Gamma_{i}=1$), consistent with \eqref{eq:ReHess}, so there the downwards gradient flow is governed by the phase piece alone. The real de~Sitter saddle $x_{i}=b_{*}$, $y_{i}=0$, with $b_{*}^{2}=-c_{i}$, is a fixed point, and linearizing about it reproduces the eigenvalues \eqref{eq:IX-eigvals} and the rays \eqref{eq:ray}. The intended experiment seeds the downwards gradient flow on the Fresnel-rotated initial surface and integrates the upwards gradient flow to verify that: every trajectory on the Lefschetz thimble contracts to $(b_{*},0)$ with no leakage to the conjugate branch $b_{-}$ or to spurious fixed points, and the flow maps the basin boundary against the Stokes locus predicted by the linearized Hessian. Reporting these integrations is deferred to future work.

\end{document}